\documentclass[twocolumn,usenatbib]{mn2e}
\usepackage{epsfig}
\usepackage{latexsym}
\usepackage{graphicx}
\usepackage{amsfonts}
\usepackage{amsmath}
\usepackage{natbib}
\usepackage{xcolor}
\usepackage{bm}
\usepackage{array}

\def\simlt{\lower.5ex\hbox{$\; \buildrel < \over \sim \;$}}
\def\simgt{\lower.5ex\hbox{$\; \buildrel > \over \sim \;$}}
\def\simgtalt{\lower.5ex\hbox{$\buildrel > \over \sim \;$}}
\def\l#1{\left#1}
\def\r#1{\right#1}

\def\bd#1{\bm{#1}}
\def\const{\hbox{{\sc const}}}

\title[Parametric component separation and CMB B-mode detection in suborbital experiments]
{Maximum likelihood, parametric component separation and CMB B-mode detection in suborbital experiments}

\author[F.~Stivoli, J.~Grain, S.~M.~Leach, M.~Tristram, C.~Baccigalupi, R.~Stompor]
{F.~Stivoli$^{1}$, J.~Grain$^{2}$, S.~M.~Leach$^{3}$, M.~Tristram$^{4}$, C.~Baccigalupi$^{3}$, R.~Stompor$^{5}$\\
$^{1}$ INRIA, Laboratoire de Recherche en Informatique, Universit\'e Paris-Sud 11, B\^atiment 490, 91405 Orsay Cedex, France
\\
$^{2}$ CNRS, Institut d'Astrophysique Spatiale, Universit\'e Paris-Sud 11, B\^atiments 120-121, 91405 Orsay Cedex, France
\\
$^{3}$ SISSA/ISAS, Astrophysics Sector, Via Beirut, 2-4, 
and INFN, Sezione di Trieste, Via Valerio 2, I-34151 Trieste, Italy
\\
$^{4}$ CNRS, Laboratoire de l'Acc\'el\'erateur Lin\'eaire, Universit\'e Paris-Sud 11, B\^atiment 200, 91898 Orsay Cedex, France
\\
$^{5}$ CNRS,  Laboratoire Astroparticule \& Cosmologie, 10 rue A. Domon et L. Duquet,
F-75205 Paris Cedex 13, France}
\begin{document}
\maketitle

\begin{abstract}

We investigate the performance of the parametric Maximum Likelihood
component separation method in the context of the CMB B-mode signal
detection and its characterization by small-scale CMB suborbital
experiments. We consider high-resolution (FWHM$=8'$) balloon-borne and
ground-based observatories mapping low dust-contrast sky areas of 400
and 1000 square degrees, in three frequency channels, 150, 250, 410
GHz, and 90, 150, 220 GHz, with sensitivity of order 1 to 10
$\mu$K per beam-size pixel.  These are chosen to be representative of some of the
proposed, next-generation, bolometric experiments.

  We study the residual foreground contributions left in the
  recovered CMB maps in the pixel and harmonic domain and discuss
  their impact on a determination of the tensor-to-scalar ratio,
  $r$. In particular, we find that the residuals derived from the
  simulated data of the considered balloon-borne observatories are
  sufficiently low not to be relevant for the B-mode science. However,
  the ground-based observatories are in need of some external
  information to permit satisfactory cleaning. We find that if such
  information is indeed available in the latter case, both the
  ground-based and balloon-borne experiments can detect the values of
  $r$ as low as $\sim 0.04$ at $95$\% confidence level.  
 The contribution of the foreground residuals to these limits is found to be then 
  subdominant
  and these are driven by the statistical uncertainty due to CMB,
  including E-to-B leakage, and noise. We emphasize that reaching such
  levels will require a sufficient control of the  level of systematic effects present
  in the data.  

\end{abstract}

\section{Introduction}
Astrophysical foregrounds are commonly recognized as one of the major
obstacles on the way to first detecting and later exploiting the
scientific potential of the Cosmic Microwave Background (CMB)
polarization signal. This is particularly the case with so called
B-mode polarization \citep{1997PhRvD..55.1830Z} due to its minute
amplitude as compared to the foreground contributions as well as CMB
total intensity and E-mode polarization signals. In fact current
foreground models \citep{2007ApJS..170..335P} generally indicate that
the foreground B-mode signal may be comparable or exceed the CMB
signal by a factor of a few in a broad range of angular scales even in
the cleanest available sky areas.  Some kind of foreground cleaning or
separation procedure will therefore be necessary and its impact on the
final `cleaned' map of the presumed CMB sky needs to be understood and
properly taken into account in its subsequent studies. Developing such
an understanding is also already of importance for the designing and
optimization of the future CMB experiments.

This has been recognized for some time and a number of studies have
been performed and published, and which have treated the problem on
different levels of generality and detail. The major challenge here is
two-fold. Firstly, there is no general recipe for propagating errors
incurred during the component separation step, i.e. for including both
the statistical uncertainty and foreground residual uncertainty.
Secondly, there is no easily calculable metric measuring the impact of
the component separation on the B-mode measurement, as both the
power spectrum or tensor-to-scalar ratio, $r$, require a proper
evaluation of the E-to-B leakage \citep{2003PhRvD..67b3501B}.

\citet{2005MNRAS.360..935T} have performed a Fisher analysis of the
problem treating the foreground residuals as an additional source of
noise, and then estimated the expected limits on $r$. In their
approach the starting point was a single 
foreground contaminated science channel and a noisy foreground
template channel from which the level of foreground residual was estimated.
Although this allows to avoid specifying in great detail a foreground
cleaning technique, no direct connection exists between the noise
values they assume and properties of any specific experiment.  They
have also neglected the impact of the E-to-B leakage. A similar
approach has been followed by \citet{2006JCAP...01..019V}, who have
attempted to link their Fisher matrix considerations to specific,
fiducial, multi-frequency data sets. The simplified error propagation
they have adopted implicitly bypasses any realistic component separation
approach, and so fails to include properly its effects on their final
results. They also neglect the presence of the E-to-B
leakage. \citet{2005PhRvD..72l3006A} performed a Fisher analysis as
well, but use specific parameters anchored in those of the
multi-frequency data set assumed.  This last work together with
\citet{2007PhRvD..75h3508A} and \citet{2009A&A...503..691B} come the
closest in the spirit to what we discuss in this paper, although
neither of the latter two works includes an actual power spectrum
estimator accounting for the leakage, what is justified at least in
part by their focus on full-sky observations.
\cite{2006MNRAS.372..615S} studied an application of an Independent
Component Analysis based approach to the component separation of
polarized partial-sky maps, resorting to the cleaned map `pseudo-spectra'
as a basis for a {\em qualitative} assessment of its performance and
relevance for the B-mode work.  \citet{2009AIPC.1141..222D} presented
a review of most of those earlier approaches, including those
incorporating a parametric approach similar to the one considered in
this work, and presented their applications in the context of a
potential future CMB B-mode satellite mission.

The approach we propose here is more specific. We focus on a
particular component separation method and power spectrum estimation
approach, which we then use to investigate the impact of the
foreground separation on the CMB B-mode detection and
characterization. The component separation method is a maximum
likelihood (ML) parametric approach~\citep{2006ApJ...641..665E} in 
a two-step implementation of
\citet{2009MNRAS.392..216S}.  The power spectrum estimator is a `pure'
pseudo-spectrum approach introduced by \citet{2006PhRvD..74h3002S}
\citep[see also][]{2007PhRvD..76d3001S} and elaborated on by
\citet{2009PhRvD..79l3515G}. Strictly speaking, our results will
therefore be specific to these two choices. However, given that these
two methods are working, promising algorithms to be implemented in the
data analysis pipelines of current and future CMB experiments, the results
should be of practical relevance for many efforts currently going on
in the field. 

We note also that whenever the frequency scaling laws can be assumed
to be nearly perfectly known, as in one of the cases we study,
and in particular in the small-sky, and thus potentially 
  statistics-starved limit, the parametric maximum likelihood (ML)
method would likely become a method of the choice, potentially
supplemented by some priors, e.g, spatial templates for all or some of
the components \citep{2009MNRAS.397.1355E}.  The results
derived here can therefore be regarded as representative and realistic
expectations for the performance of classes of
the future experiments we consider. Moreover, 
part of the analysis presented here can be straightforwardly applied
to any component separation method in which foreground
  spectral and amplitude parameters are estimated in separate steps.

Our focus in this work is on suborbital experiments. Those have a
potential advantage of selecting the cleanest sky areas, but suffer
due to the cut-sky effects.  They also usually have a limited number
of frequency channels with which to observe the sky.  We consider two
kinds of experiments: those with an access to the high frequencies
($\simgtalt 250$ GHz) referred to as balloon-borne, and those
with access limited to frequencies lower than $250$ GHz, referred to
as ground-based. We will also consider some combination and extensions
of these two cases.  We then apply our proposed analysis chain
to simulated data for different foreground case studies,
allowing for different levels of mismatch between the assumptions made
on the analysis and simulation stages, in order to evaluate the impact
of the component separation residuals first on the recovered B-mode
power spectrum and later on the value of a $r$ which can be derived
from such data.

The paper is organized as follows. In Sections~\ref{sec:miramare} and
\ref{sec:xpure}, we first provide brief descriptions of the specific
data analysis techniques and their implementation, used throughout
this paper. In Section~\ref{sec:mockData} we describe our
  simulated sky model, and in Section~\ref{sec:mock} we define the
  experimental characteristics and a set of foreground case studies.
Our results are presented in Section~\ref{sec:Results}, and
their analysis, concerning the residuals and their impact on the
cosmological B-mode detection, is given in Sections~\ref{sect:anaRes}
and~\ref{sec:tensor2scalar}, respectively.

\section{Parametric component separation method}
\label{sec:miramare}
In this section we briefly outline the parametric component separation
algorithm proposed by \citet{2009MNRAS.392..216S}. The multi-frequency
sky signal is modelled as
\begin{eqnarray}
\bd{d}_p = \bd{A}_p\,\bd{s}_p\,+\,\bd{n}_{p},
\label{eqn:dataModelIdeal}
\end{eqnarray}
where $\bd{d}_p$ is a vector containing the data from $N_{freq}$
frequencies assumed to share a common angular resolution, $\bd{s}_p$
is a vector of $N_{comp}$ signal amplitudes to be estimated, $\bd{A}_p
\equiv\bd{A}_p\l(\bd{\beta}\r)$ is a component `mixing' or frequency
scaling matrix with a total of $N_{spec}$ free `spectral parameters'
$\bd{\beta}$ also to be estimated, and $\bd{n}_{p}$ is the noise at
each pixel $p$.  We can write down a likelihood for the data of the
form
\begin{eqnarray}
-2\,\ln\,{\cal L}_{data}\l(\bd{s}, \bd{\beta}\r) = \const +
\l(\bd{d}-\bd{A}\,\bd{s}\r)^t\,\bd{N}^{-1}\l(\bd{d}-\bd{A}\,\bd{s}\r),
\label{eqn:genMpixLike}
\end{eqnarray}
where $\bd{N}$ is the noise covariance matrix of the data and we have
now dropped the pixel index $p$. This likelihood reaches its maximum
for the values of $\bd{s}$ and $\bd{\beta}$ fulfilling the relations,
 \begin{eqnarray}
-\l(\bd{A}_{,\bd{\beta}}\,\bd{s}\r)^t\,\bd{N}^{-1}\,\l(\bd{d}-\bd{A}\,\bd{s}\r) = 0
\label{eqn:maxRels}
\\
\bd{s} = \l(\bd{A}^t\,\bd{N}^{-1}\,\bd{A}\r)^{-1}\,\bd{A}^t\,\bd{N}^{-1}\,\bd{d},
\label{eqn:step2amps}
 \end{eqnarray}
where $_{,\bd{\beta}}$ denotes a partial derivative with respect to
$\bd{\beta}_i$. Under the assumption that the spectral parameters are
the same for a collection of pixels, corresponding to the physical
assumption that the spectral parameters vary more slowly in space than
the signal amplitudes, it is possible to substitute the generalised
least squares solution Equation~(\ref{eqn:step2amps}) into the
likelihood Equation~(\ref{eqn:genMpixLike}), thereby eliminating the
sky signals $\bd{s}$, in order to obtain a spectral index likelihood
given by
\begin{eqnarray}
-2\ln {\cal L}_{spec}\l(\bd{\beta}\r) &= & \hbox{{\sc const}}
\label{eqn:slopelikeMpix}
\\
&-&\l(\bd{A}^t\,\bd{N}^{-1}\,\bd{d}\r)^t\,\l(\bd{A}^t\,\bd{N}^{-1}\,\bd{A}\r)^{-1} \l( \bd{A}^t\,\bd{N}^{-1}\,\bd{d}\r).
\nonumber
\end{eqnarray}
The spectral parameters that minimize
Equation~(\ref{eqn:slopelikeMpix}) can be found using numerical
techniques, and then substituted into Equation~(\ref{eqn:step2amps})
in order to find the corresponding signal amplitudes pixel by
pixel. Finally, the noise covariance matrix, describing the properties of the
noise contained in the data $\bd{d}$, is propagated to the component
estimates $\bd{s}$: 
\begin{eqnarray}
\bd{N_s} \equiv \l(\bd{A}^t\,\bd{N}^{-1}\,\bd{A}\r)^{-1}.
\label{eqn:noiseCorrOptim}
\end{eqnarray}

{\sc Miramare}\footnote{\tt people.sissa.it/$\sim$leach/miramare},
which is our implementation of this two-step algorithm, uses codes
from the {\sc CosmoMC} package \citep{2002PhRvD..66j3511L} in order to
perform an initial conjugate gradient descent to the minimum of the
spectral index likelihood Equation~(\ref{eqn:slopelikeMpix}). This is
followed by estimation of the curvature of the spectral index
likelihood on a regular grid, which then forms the basis of the
`proposal function' for drawing spectral index samples using the
Markov Chain Monte-Carlo (MCMC) technique. Once the spectral
parameters have been determined from an analysis of the MCMC samples,
these values are substituted into Equations~(\ref{eqn:step2amps}) and
(\ref{eqn:noiseCorrOptim}) in order to obtain the signal amplitudes
and their covariance.

We can check our assumption of the constancy of the spectral
parameters and the overall `goodness of fit' by evaluating the
log-likelihood Equation~(\ref{eqn:genMpixLike}) at the maximum
likelihood value, and comparing it with the number of degrees of
freedom given by
\begin{eqnarray}
N_{dof}=  N_{pix}\times (N_{freq} - N_{comp}).
\label{eqn:dof}
\end{eqnarray}
A poor fit, for instance, due to either
the wrong parametrization assumed for the components present in the 
data or the spatially variability of the parameters, will be accompanied by an excessive log-likelihood.
Other goodness of fit tests are also possible and will be discussed
elsewhere.

The ML formalism allows to straightforwardly incorporate the uncertainty 
due to errors
on the calibration of the input maps. This can be done by replacing the mixing
matrix, $\bd{A}$, by a product of a diagonal matrix $\bd{C}$, representing the
calibration for each of the single channel maps, and the foreground mixing
matrix, $\bd{A}$. Following \citet{2009MNRAS.392..216S}, we denote
the diagonal elements of $\bd{C}$ as, $\omega_i$, so,
\begin{eqnarray}
\bd{C} = {\rm diag}\l(\omega_i\r)_{i=0,\dots, n_f-1}.
\end{eqnarray}
Such a problem is clearly degenerate if no external constraints are
imposed on the calibration parameters.  In a case of actual
observations, prior information on the calibration uncertainty is in
fact usually available and can be often expressed as a Gaussian error
centered around some most likely value. Mathematically, it just
corresponds to multiplying the likelihood in
Equation~(\ref{eqn:slopelikeMpix}) by the relevant priors.

We note that if the absolute calibration of the resulting component
maps is not required one could in principle reduce the number of
calibration parameters by one by subsuming one of the $\omega_i$, say
that of the very first channel, $\omega_0$, factors into the overall
normalization of the sought-after component maps. This is the
  approach that we will use later in this paper, always assuming that
  the uncertainty of the relative calibration of the higher frequency
  channels with respect to the lowest frequency channel can be
  sufficiently well described as a Gaussian random variable with a
  known width.

\section{Polarized power spectrum estimation}

\label{sec:xpure}

In this section we briefly describe the approach that we use to
measure the E- and B-mode power spectra which is based on the `pure'
pseudo-$C_\ell$'s method proposed in
\cite{2006PhRvD..74h3002S}. Further details of the implementation and
tests on simulations can be found in \cite{2006PhRvD..74h3002S},
\cite{2007PhRvD..76d3001S} and in \cite{2009PhRvD..79l3515G}. The
latter work also extends the pure approach to the case of
cross-spectra. The pure pseudo-spectrum estimators retain speed and
efficiency of the standard pseudo-spectrum
methods and have been devised to suppress the effect of
the E-to-B leakage, thus ensuring the near optimality of the
estimated B-mode power spectrum. The performance of such an estimator 
is demonstrated in the last part of this section where the 
{\it statistical} uncertainties on the E- and B-mode reconstruction
is evaluated thanks to Monte-Carlo simulation. 

\subsection{The pure pseudo-$C_\ell$'s estimator}
In general a polarization field on the partial sky can be divided into
three classes of modes: pure E-modes, pure B-modes and
ambiguous modes, which are a mixture of the true E and B
modes~\citep{2003PhRvD..67b3501B}.  The standard pseudo-spectrum
approach consists of projecting the polarization fields
\begin{eqnarray}
\mathbf{P}=\left(\begin{array}{c}
		Q \\
		U
	\end{array}\right)
\end{eqnarray}
onto the full-sky harmonic basis of B-modes 
\begin{eqnarray}
\mathbf{Y}^B_{\ell m}=\frac{1}{2}\sqrt{\frac{(\ell-2)!}{(\ell+2)!}}\left(\begin{array}{c}
		i(\partial^2-\bar\partial^2) \\
		\partial^2+\bar\partial^2
	\end{array}\right)Y_{\ell m},
\end{eqnarray}
where $\partial\,(\bar\partial)$ stands for the spin-raising (lowering)
operator. However, this basis contains both pure B-modes and ambiguous
modes on the partial sky.  Consequently the standard pseudo-$C_\ell$
estimator includes these ambiguous modes in the B-mode power spectra
estimates, thus resulting in contamination from the much larger
E-mode contribution -- an effect hereafter referred to as E-to-B
leakage. This can be removed on averaged, but will still lead to a
significant increase of the estimated spectrum variance.

The ambiguous modes can however be filtered out by projecting the
polarization field onto a pure B-mode basis. Such a basis is
constructed from the spherical harmonics and a particular window
function $W$, such that $W$ and $\partial{W}$ vanish on the boundary
of the observed sky:
\begin{eqnarray}
\mathcal{Y}^{B}_{\ell m}=\sqrt{\frac{(\ell-2)!}{(\ell+2)!}}\left(\begin{array}{c}
		i(\partial^2-\bar\partial^2) \\
		\partial^2+\bar\partial^2
	\end{array}\right)WY_{\ell m}.
\label{eqn:pureBasis}
\end{eqnarray}
Pseudo-multipoles free of E-to-B leakage can be then computed from this
basis by taking the dot product
\begin{eqnarray}
	\tilde{a}^B_{\ell m}=\displaystyle\int d\vec{n}~\mathcal{Y}^{B~\dag}_{\ell m}\cdot \mathbf{P},
\end{eqnarray}
from which a pseudo-power spectra can be derived
\begin{eqnarray}
	\tilde{C}^B_\ell=\frac{1}{2\ell+1}\displaystyle\sum_m\tilde{a}^B_{\ell m}\tilde{a}^{B\star}_{\ell m}.
\end{eqnarray}
The pseudo-power spectrum for the E-modes is identically derived
by projecting the Stokes parameters onto the harmonic basis for the E-modes,
defined as:
\begin{eqnarray}
\mathcal{Y}^{E}_{\ell m}=\frac{1}{2}\sqrt{\frac{(\ell-2)!}{(\ell+2)!}}\left(\begin{array}{c}
		\partial^2+\bar\partial^2 \\
		-i(\partial^2-\bar\partial^2)
	\end{array}\right)WY_{\ell m}.
\end{eqnarray}
Unbiased estimates for both the E- and B-mode power spectra are finally obtained by solving the linear system
\begin{equation}
	\left(\begin{array}{cc}
			M^{(+)}_{\ell\ell'} & M^{(-)}_{\ell\ell'} \\
			M^{(-)}_{\ell\ell'} & M^{(+)}_{\ell\ell'}
		\end{array}\right)\left(\begin{array}{c}
			C^E_{\ell'} \\
			C^B_{\ell'}
		\end{array}\right)=\left(\begin{array}{c}
			\tilde{C}^E_{\ell}-N^E_{\ell} \\
			\tilde{C}^B_{\ell}-N^B_\ell
		\end{array}\right)
\end{equation}
where $N^{E/B}_{\ell'}$ stands for the noise bias and
$\mathbf{M}^{(+/-)}_{\ell\ell'}$ for the different mode-mode coupling
matrices. We emphasize that the pure formalism adopted hereafter
corrects for the E-to-B leakage due to partial sky coverage only,
leading to elements of $M^{(-)}$ much smaller than those of
$M^{(+)}$. However, such off-diagonal elements are not strictly zero
because of pixel-induced E-to-B leakage. This remaining leakage
between the two type of polarization is nevertheless very small and
carefully taken into account in the computation of the mode-mode
coupling matrices.

The required extensions are included in our implementation of the pure
pseudo-$C_\ell$'s estimator, called {\sc Xpure}, which we use in this work. 
{\sc Xpure}  is a generalization of the {\sc Xspect} and {\sc Xpol} codes
\citep{2005MNRAS.358..833T}. It can handle multiple maps for computing
auto- and cross-power spectra. The different mode-mode
couplings due to partial sky coverage (i.e. $\ell$-to-$\ell'$ mixing) and pixelization (i.e.residual E-to-B leakage) are accounted for during the
mode-mode coupling matrix computation. The code is based on
the {\sc S$^2$HAT} library\footnote{{\sc S$^2$HAT}: {\tt www.apc.univ-paris7.fr/$\sim$radek/s2hat.html}\\
  {\sc pureS$^2$HAT}: {\tt www.apc.univ-paris7.fr/$\sim$radek/pureS2HAT.html}} -- a parallel
library allowing for efficient computation of spin-weighted
spherical harmonic transforms.
\begin{figure*}
\begin{center}
  \includegraphics[width=7.5cm]{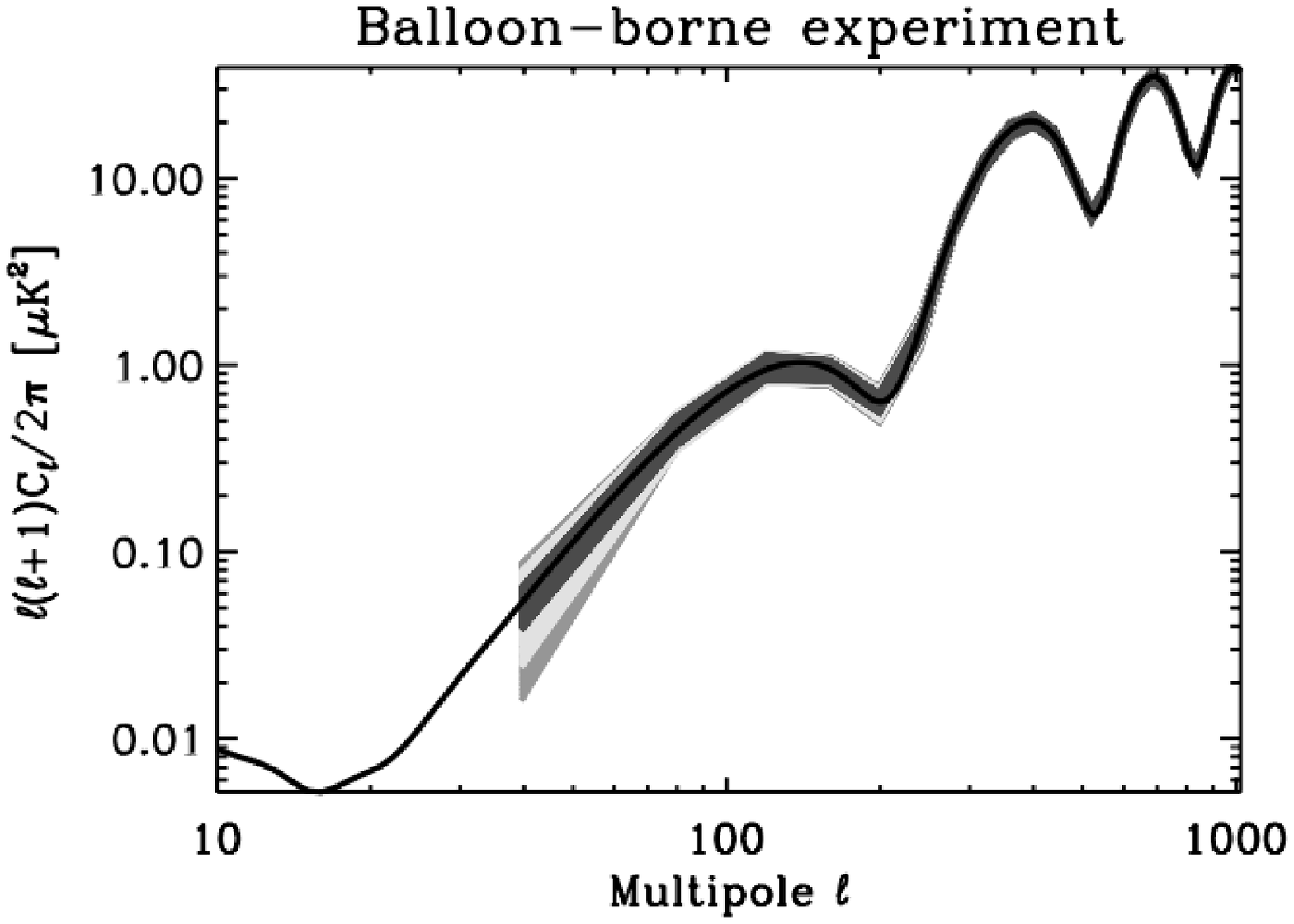} \includegraphics[width=7.5cm]{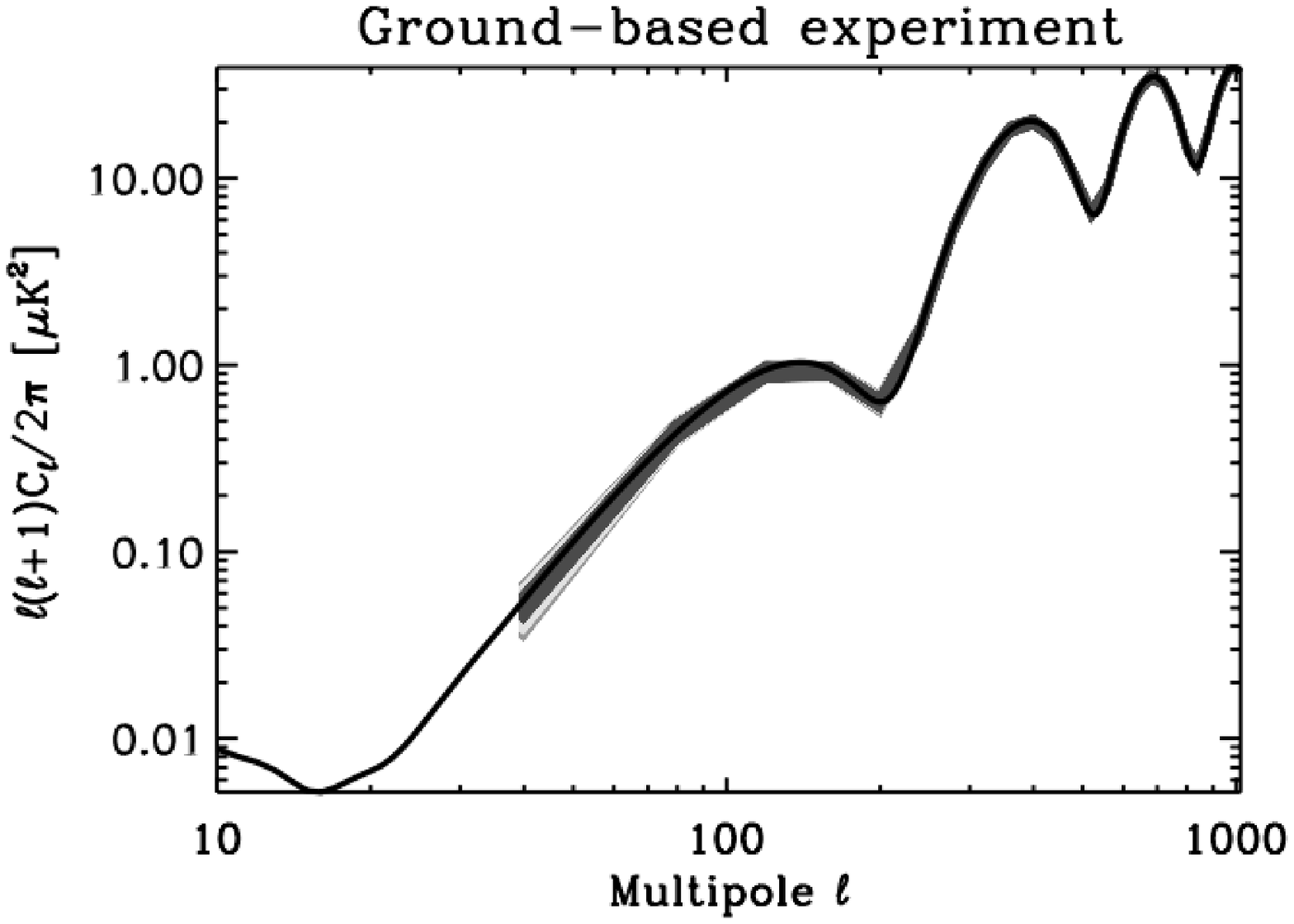}
  \includegraphics[width=7.5cm]{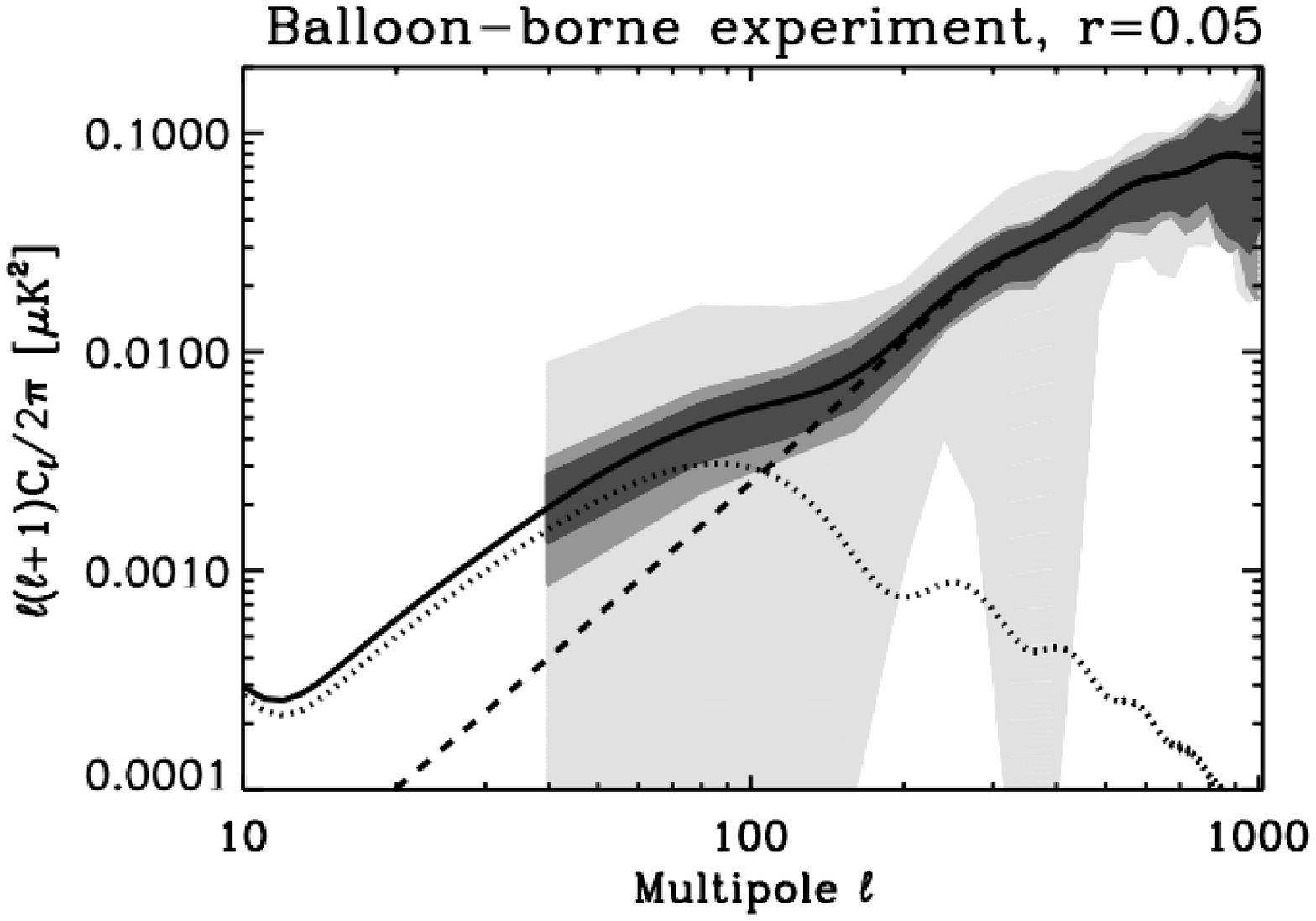} \includegraphics[width=7.5cm]{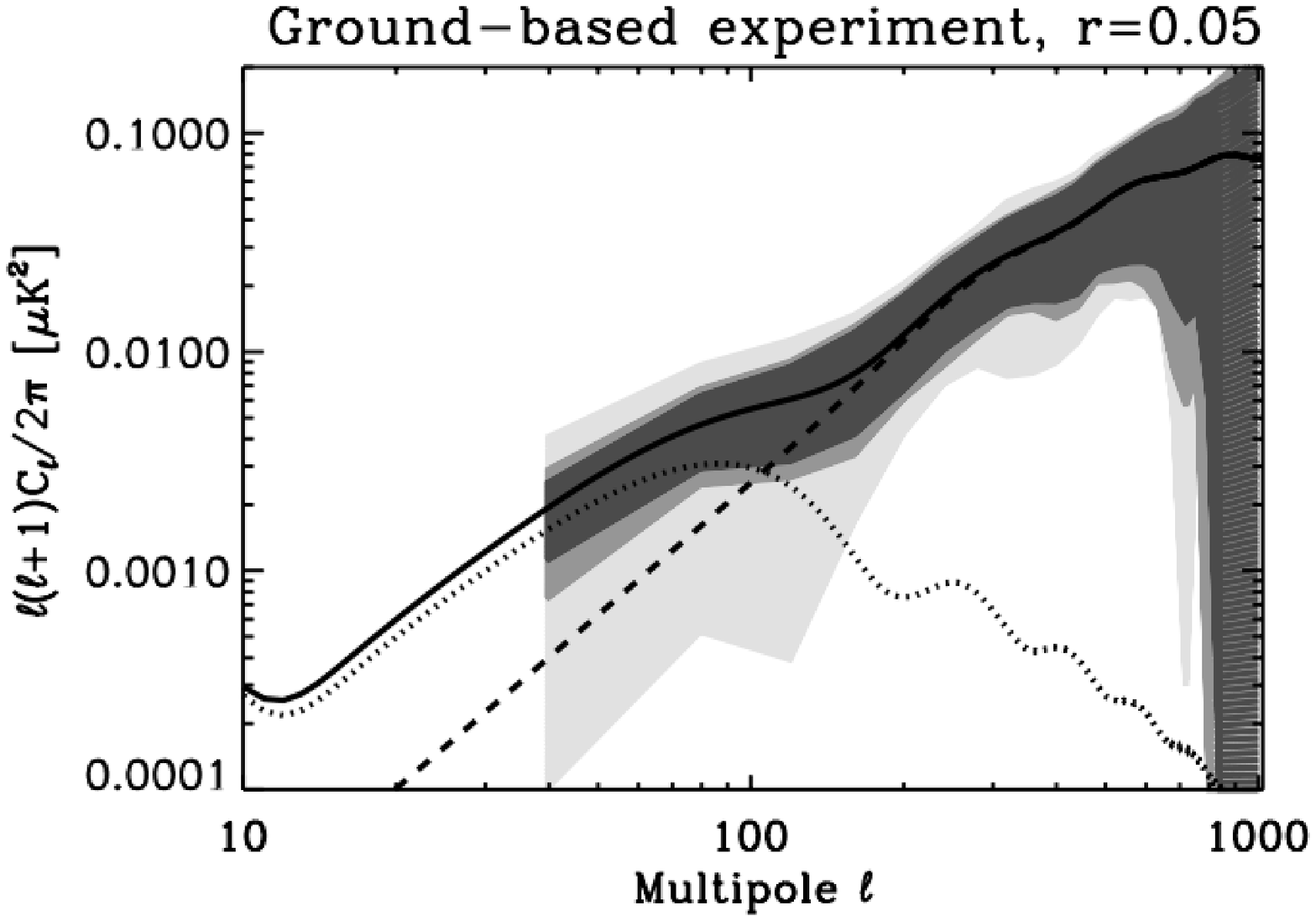}
 \caption{ Variance of the estimated E-mode (upper panels) and B-mode
 power spectrum (lower panels) for the balloon-borne (left panels) and
 ground-based (right panels) sky-coverage as shown in
 Figure~\ref{fig:masks}. From darkest to lightest grey: Fisher
 estimation (i.e. the so-called $f_{\rm sky}$ approximation), standard
 deviation of 500 Monte-Carlo simulations using the pure
 pseudo-$C_\ell$'s estimator as implemented in {\sc Xpure} and,
 standard deviation of 500 Monte-Carlo simulations using the standard
 pseudo-$C_\ell$'s estimator as implemented in {\sc Xpol}. We stress
 that {\sc Xpure} provides lower variances than {\sc Xpol} for the
 B-modes, while {\sc Xpol} performs better for the E-modes.}
  \label{fig:specExample}
\end{center}
\end{figure*}

\subsection{Sky apodization}
\label{Sec:Apodize}
The applicability of the pure pseudo-$C_\ell$'s estimator strongly
depends on being able to compute the sky apodizations which are needed
in the calculation of the relevant pure basis functions,
Equation~(\ref{eqn:pureBasis}), and which have to fulfill appropriate
boundary conditions. Different functions have been proposed, ranging
from those derived from an optimization procedure to those based on
some analytic expressions. Their relative merit has been extensively
discussed by \cite{2009PhRvD..79l3515G} who showed that the optimized
sky apodization scheme proposed in \cite{2007PhRvD..76d3001S} leads to
the lowest variance on the power spectrum estimation.  

The underlying strategy for deriving such an optimized sky apodization
is to search for the $W$ functions which make the pure pseudo-$C_\ell$
as close as possible to optimal, quadratic power-spectrum estimator
\citep[see Section V of][]{2007PhRvD..76d3001S}. An optimized
weighting scheme therefore involves a specific sky apodization for
each $\ell$-band for which the power spectrum is to be estimated,
according to the signal and noise power aliasing in each band. As
shown in \citep{2007PhRvD..76d3001S}, this optimization procedure can
deal with both external and internal boundaries, due to limited sky
coverage and holes induced by point source removal, respectively. Such an
optimization procedure however assumes that noise and signal are
known. Although this is a good assumption for noise and E-modes (which
can be recovered precisely enough without optimization), this does not
hold for the B-modes and an erroneous prior may introduce
suboptimality in the B-modes estimate. However, it has been shown by
\cite{2009PhRvD..79l3515G} that the B-modes prior only mildly affects
the performance of the power-spectrum estimation using the optimized
sky apodization. This weak dependence is in fact a direct consequence
of the optimization process, which attempts to select an apodization
to reduce first the sampling variance due to the E-to-B leakage and
second the noise variance, with the B-mode variance itself usually
being subdominant.

These optimized sky apodizations, leading to the lowest statistical uncertainties, will be used throughout this work to 
compute the polarized power spectra from the CMB maps estimated from the
component separation process.

\subsection{Statistical uncertainties}
\label{sec:StatVariance}
A complete characterization of the error budget incurred by both the
foreground cleaning and power spectrum estimation processes is
mandatory for setting statistically meaningful constraints on the
tensor-to-scalar ratio. We evaluate the sampling and noise variance
induced by the power spectrum estimation stage using Monte-Carlo (MC)
simulation, allowing us to determine the statistical part of the total
error budget and to demonstrate the performance of the pure
pseudo-$C_\ell$'s estimator.

\begin{figure}
\begin{center}
  \includegraphics[width=6cm]{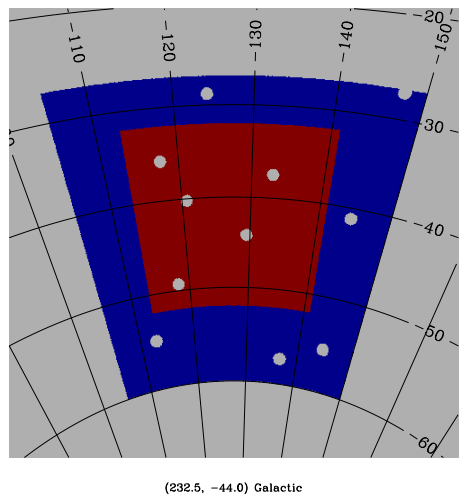}
  \caption{Sky areas for the balloon-borne (inner) and
      ground-based (outer, larger patch) experimental setups.  The
      holes mimic the effect of masking resolved point sources and are
      included in our optimized apodization calculation.}
  \label{fig:masks}
\end{center}
\end{figure}

The observed sky area for the balloon-borne and ground-based
experiments considered hereafter are shown in
Figure~\ref{fig:masks} where holes due to point-source removal are
carefully taken into account. The sky fraction is roughly
{\rm $f_{\rm sky}=1\%$} and 2.5\% for the balloon-borne and ground-based
experiments, respectively. We assume homogeneous, white noise which
is deduced from the noise per frequency channels using
Equation~(\ref{eqn:noiseCorrOptim}). This gives a noise level in the CMB
maps of approximately 1.65$\mu$K and 3.2$\mu$K per $3.5'$ pixel for the
balloon-borne and the ground-based experiments respectively. The
simulations use the WMAP 5-year best-fit cosmological
model~\citep{2009ApJS..180..306D} and the simulated B-mode includes
the lensing and primordial component with a tensor-to-scalar ratio,
$r$, equal to 0.05.  The E- and B-mode power spectra are estimated by
downweighting the simulated maps with the optimized sky apodization as
described in the previous section and finally binned with a band width
of $\Delta\ell=40$, with the lowest $\ell$-bin starting at $\ell=20$.

We demonstrate the performance of the pure estimator
in Figure~\ref{fig:specExample} which shows the input E- and B-mode
power spectra and the estimated variances of the reconstructed E- and
B-mode power spectra derived as the standard deviation from 500
MC simulations. The three variances displayed on this
figure are obtained from, 
i) the Fisher estimate, or so-called $f_{\rm sky}$-formula, ii) the
standard deviation of 500 MC simulations using pure pseudo-$C_\ell$'s
estimator and iii) the standard deviation of 500 MC simulations using
standard pseudo-$C_\ell$'s estimator. 

For the B-modes, the variance is significantly reduced by using the
pure pseudo-$C_\ell$'s estimator at angular scales where sampling
variance is dominating ($\ell\leq700$), while the two approaches lead
to similar variances at smaller angular scales where noise is the
dominant contribution to the statistical uncertainties. More
specifically, this technique is mandatory for the balloon-borne
experiment to be able to extract the B-mode from the maps of the two
Stokes parameters at intermediate and large angular scales
 ($\ell\leq400$) while the variance is reduced by a factor 2 at
those scales for the ground-based experiment. Moreover, the pure
pseudo-$C_\ell$'s estimator is required for both type of experiments
to be able to {\it statistically} disentangle the inflationary 
gravitational waves at
$\ell\leq100$ from the secondary, lensing-induced B-mode.

For the E-modes, the standard pseudo-$C_\ell$'s estimator is
preferred, for achieving higher accuracy (at least at large angular
scales for the balloon-borne sky coverage). The reason for such a
higher efficiency of the standard approach as compared to the pure one
for E-mode is two-fold: on the one hand, B-to-E leakage only mildly
affects the variance in the standard approach and, on the other hand,
ambiguous modes are mainly composed of E-modes, and so a significant
amount of information may be lost by removing them using the pure
approach.

\section{Simulated sky}

\label{sec:mockData}

In this section we describe the sky model we use for simulations in
this work. We introduce the `basic model', which is a simple
model of the sky signal consistent with the currently available
information on diffuse foregrounds, and then discuss a number of
simple extensions whose physical parameters are poorly constrained
by current observations.

\subsection{Sky signals - Basic model}
\label{sec:simulated_sky}

Our analysis area is centered around RA$=62^{\circ}$ and
Dec$=-45^{\circ}$. This sky region, which is in the anti-sun direction
during the austral summer, has already been observed by several CMB
polarization experiments including
Boomerang~\citep{2006ApJ...647..813M},
ACBAR~\citep{2009ApJ...694.1200R}, QUAD~\citep{2009ApJ...705..978B}
and will be targeted by future experiments like EBEX
\citep{2004SPIE.5543..320O, 2008SPIE.7020E..68G}.  We first note that
WMAP-based estimates of the level of unresolved point source power
together with conservative assumptions about the expected level of
radio source polarization suggest that this signal will be negligible
compared to the lensing B-mode signal (Ben Gold, private
communication). Similar conclusions hold for the infra-red sources.
Certainly though, a few bright extragalactic sources
will be present in the field both by chance and for the purposes of
in-flight beam mapping and calibration. The process of masking out
these sources is mimicked in our simulations, similar to what done in
this respect by \citet{2007PhRvD..76d3001S}, as shown in
Figure~\ref{fig:masks}.

Our Galactic sky model on this relatively high Galactic latitude
region consists of two diffuse foreground components: synchrotron and
thermal dust emission.  The synchrotron total intensity emission was
simulated using the 408 MHz map of \citet{1982S&T....63..230H}, with
free-free emission subtracted and small-scale power added by
\citet{2002A&A...387...82G}. We extrapolated this
template\footnote{{\tt ftp.rssd.esa.int/pub/synchrotron}} up to $65$
GHz, using the WMAP five-year maximum entropy method derived
synchrotron spectral index map~\citep{2009ApJS..180..265G}

The thermal dust total intensity emission is taken from the combined
COBE-DIRBE and IRAS dust template of \cite{1998ApJ...500..525S},
extrapolated in frequency using the scaling accounting for temperature
variations as described in \citet{1999ApJ...524..867F}. We first
extrapolate the dust down to $65$ GHz for matching our polarization
model with WMAP data as indicated below, and then extrapolate it back
into our chosen frequency range according to a uniform greybody
scaling law inspired by FDS Model 3 \citep{1999ApJ...524..867F}
\begin{eqnarray}
\label{eq:dust_scal}
A_{\rm dust}\propto\frac{{\nu}^{\beta_{d}+1}}{\exp \frac {h\nu}{kT_d}-1}\ ,
\end{eqnarray}
where $T_d=18.0$K and $\beta_d=1.65$.

In order to simulate the polarization in this region, we normalized
the large-scale polarization amplitude to the E and B spectra
estimated by \citet{2007ApJS..170..335P}. This normalization is
achieved by first assuming that the polarized intensity of both
synchrotron and dust is proportional to the total intensity, which
introduces two free parameters, $p_{\rm dust}$ and $p_{\rm sync}$. In
order to obtain the large-scale polarization angles $\theta$, we take the
$Q$ and $U$ templates from the WMAP polarized dust template, which is
based on information derived from starlight polarization
and a geometric suppression factor taken from a three dimensional Galactic
magnetic field model described in \cite{2007ApJS..170..335P}. This template
introduces a large-scale modulation to the polarization pattern of the
synchrotron and dust.  We then add Gaussian fluctuations to the
polarization angles on smaller scales following the method of
\citet{2002A&A...387...82G}, assuming a model $C_{l}^{\cos
2\theta,\sin 2\theta}\propto l^{\alpha}$, where $\alpha=-3$.
\begin{figure*}
\centering
\includegraphics[width=4cm]{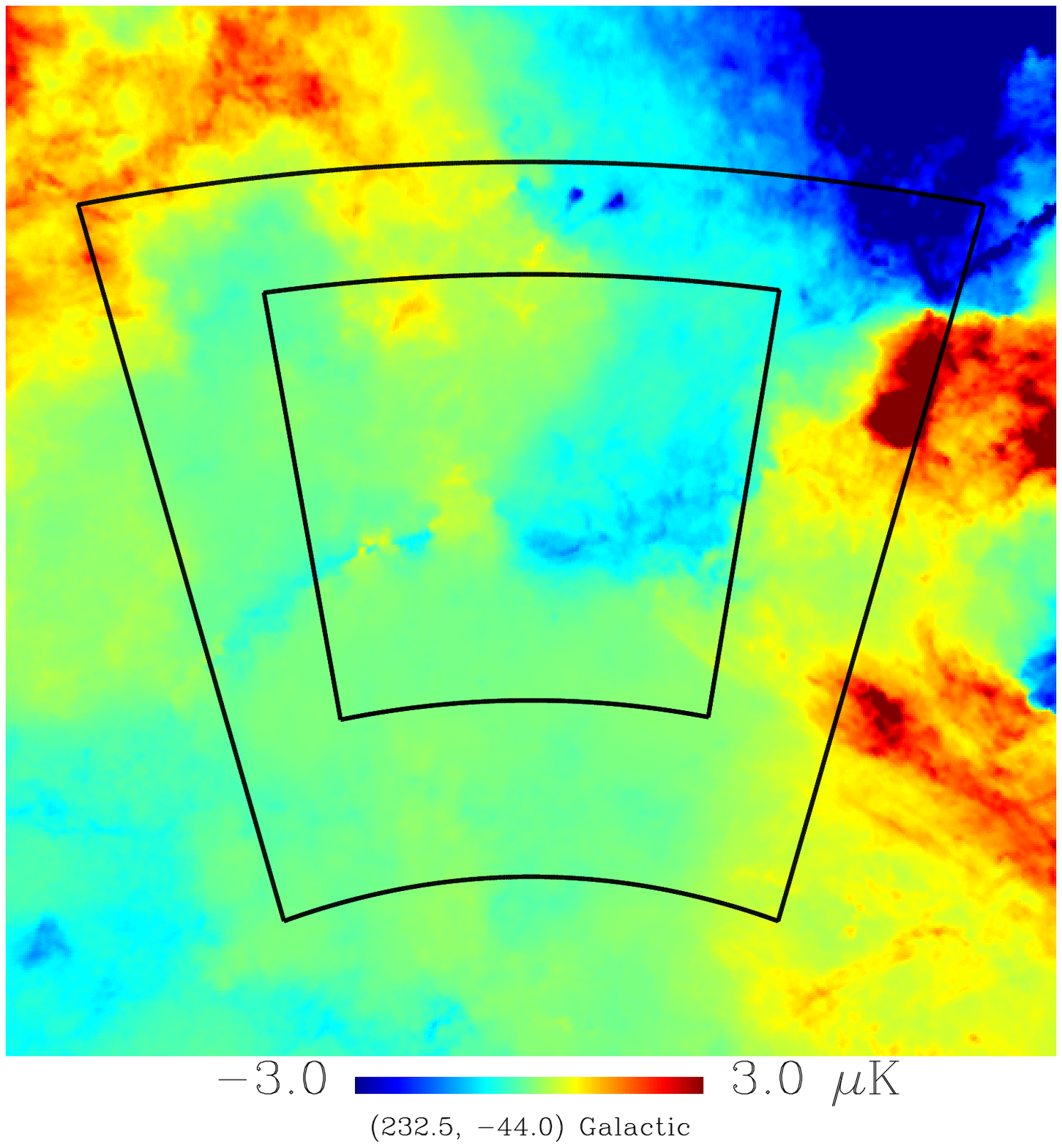}
\includegraphics[width=4cm]{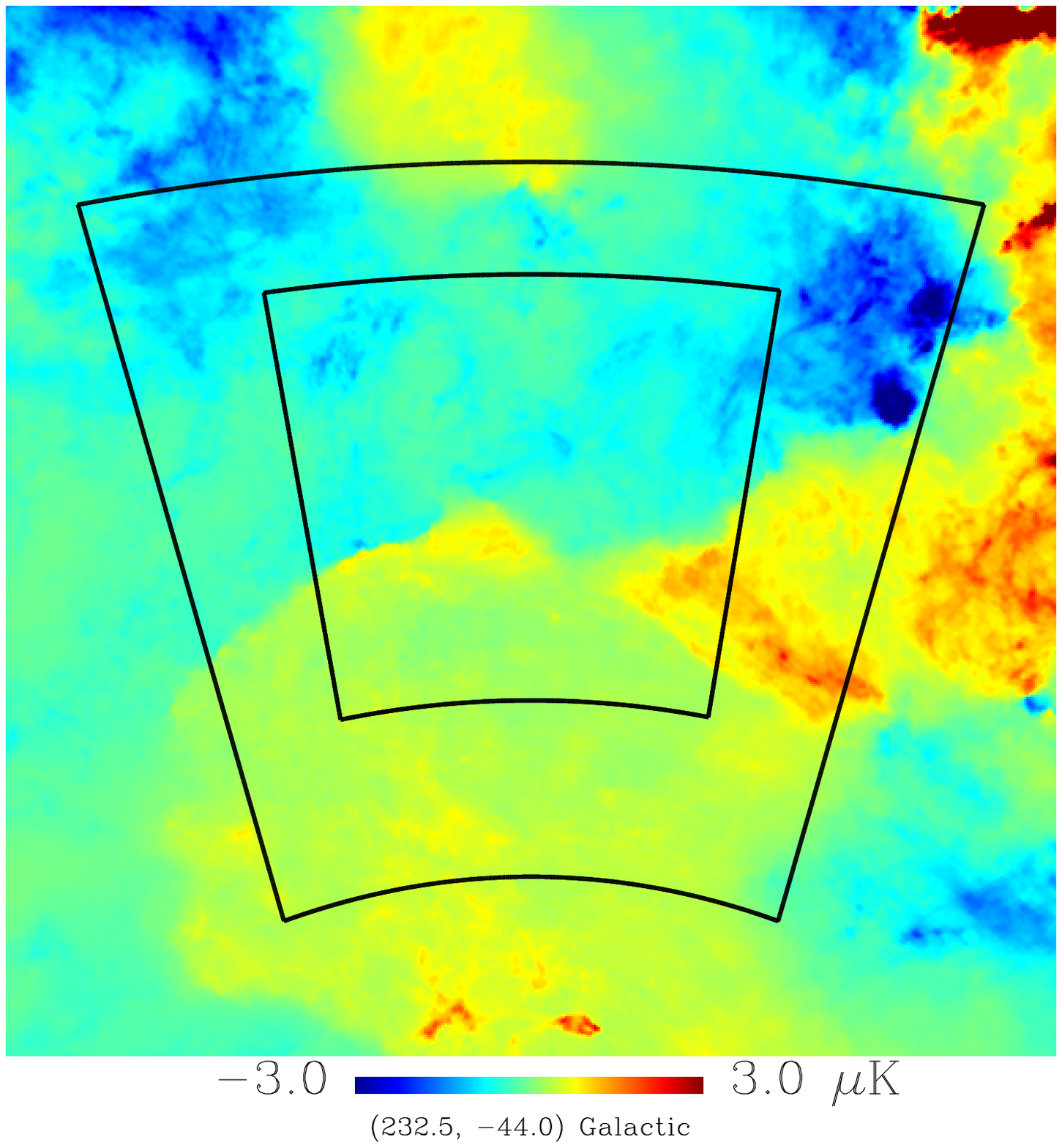}
\includegraphics[width=4cm]{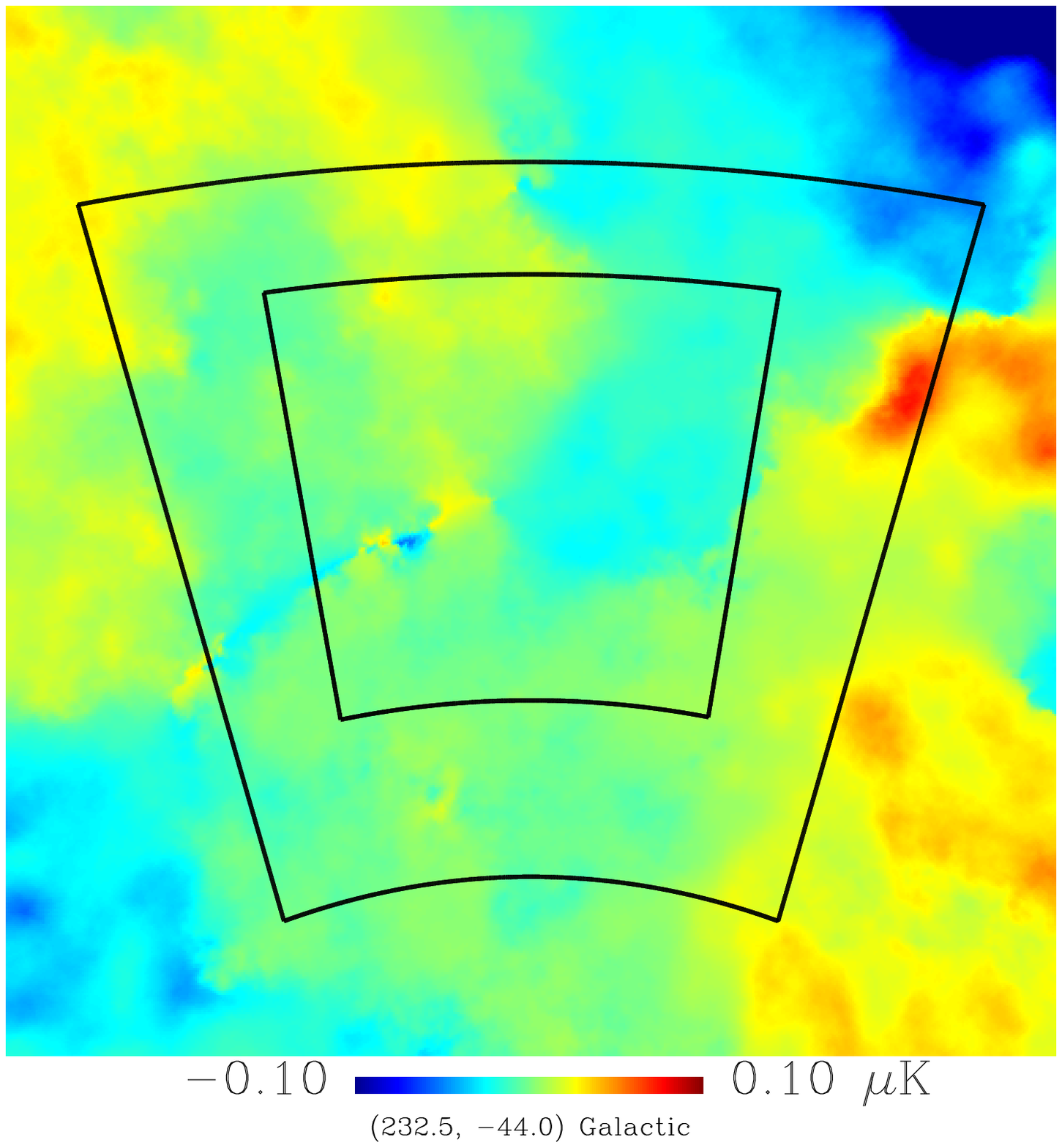}
\includegraphics[width=4cm]{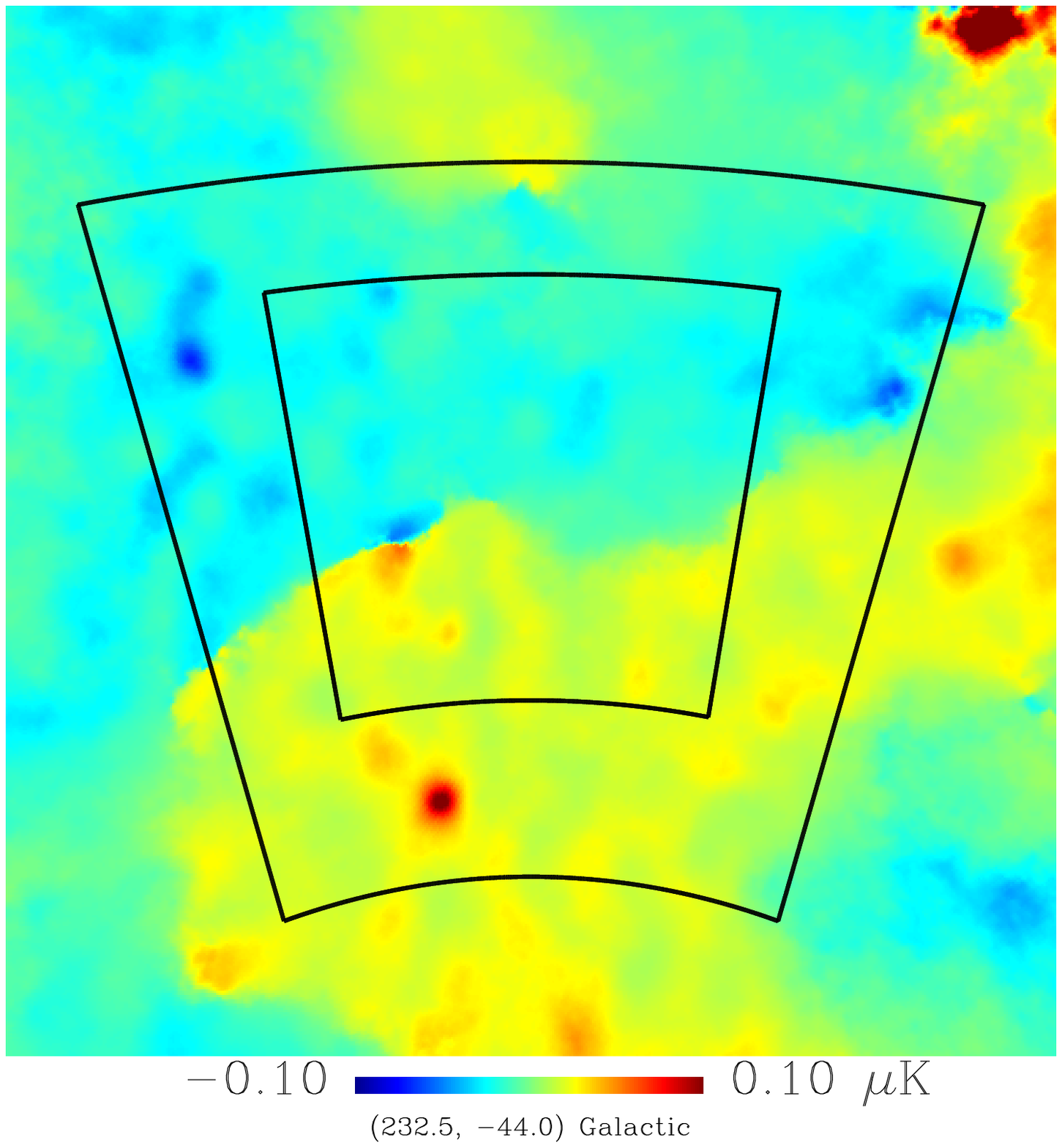}
\caption{Simulated thermal dust (left) and synchrotron (right
panels) Stokes Q and U templates at 150 GHz, for the `basic
model' described in Section~\ref{sec:mockData}. The high
degree of correlation between the two components on large angular
scales is imposed by polarization angle template we assume. The two
foregrounds have roughly the same amplitude around 70 GHz.}
\label{sky_template}
\end{figure*}

With these templates of $I$, $Q$ and $U$, we evaluate the power
spectra for E and B modes in order to finally normalize the $Q$ and
$U$ maps to an effective polarization fraction. A good match is found
for $p_{\rm dust}=p_{\rm sync}=0.1$.  In Figure~\ref{sky_template} we
show maps of the synchrotron and dust templates at $150$ GHz. The
large-scale modulation introduced by the WMAP polarized dust template
is clear from the correlated appearance of the synchrotron and dust. A
certain amount of correlation between these polarized components is
expected because the Galactic magnetic field (GMF) is a key ingredient
common to both the physics of synchrotron emission and dust
alignment. Indeed theoretical modelling of the GMF is underway by
several groups
\citep{2007ApJS..170..335P,2008A&A...490.1093M,2009A&A...495..697W,2010MNRAS.401.1013J}. However,
a global model of both synchrotron and dust polarization in a
turbulent GMF at high galactic latitudes is currently unavailable 
(see however the recent work by \citet{2010arXiv1003.4450F}), and
this is why we adopt the semi-empirical approach to our GMF
simulations just described. Such a modeling can potentially
exaggerate the overall correlation level between these two components
by extending it to smaller angular scales.  This can in turn have
important consequences for the performance of the component separation
process, as discussed in Section~\ref{subsect:groundBasedCases}.

In the range of frequencies relevant here, the synchrotron
  contribution is subdominant compared to the dust but
  increases monotonically with decreasing frequency.  These two
  components become comparable at around 70 GHz, where the total
  foreground minimum is also found.

\subsection{Sky signals - Extensions}
The problem of insufficient frequency coverage is exacerbated by
effects that can generally be described as foreground complexity,
which will increase the biases in the recovered components. We now
introduce a few extensions to the basic model just described that
we will investigate later in our analysis. 

\subsubsection{Extra small-scale power}
\label{sec:extra_power}

We investigate how increasing the foreground power on small angular
scales impacts on foreground cleaning by varying the parameter
$\alpha$. While in the basic model, the small-scale power in
polarization is rapidly decaying with $\alpha=-3$, we also study a
more extreme case with $\alpha=-2$.

\subsubsection{Spatially-varying frequency scaling}
\label{sec:varying_spectral_index}
\begin{figure}
\centering
\includegraphics[width=6cm]{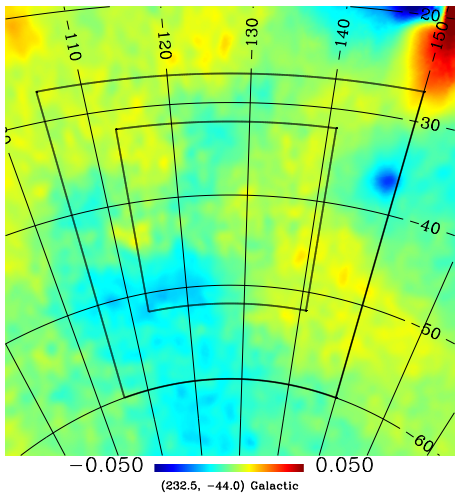}
\caption{The dust spectral index variations at 150 GHz adopted
  in this paper \citep{1998ApJ...500..525S}.}
\label{effective_spectral_index}
\end{figure}

Foreground spectral index spatial variations, if poorly estimated,
will compromise the estimation of the CMB signal. In turn this
compromises the measurement of the CMB power spectrum, particularly on
the angular scales on which the spectral index varies.  In the basic
model just described the dust scales uniformly as
Equation~(\ref{eq:dust_scal}). We will also investigate the impact of
the FDS temperature variations~\citep{1998ApJ...500..525S}, whose
effective powerlaw spectral index (minus the mean) at 150 GHz is
shown in Figure~\ref{effective_spectral_index}, and whose RMS
  variation is 0.008.

Approaches for dealing with spectral index variations include Taylor
expanding the foreground continuum emission and fitting the spectral
index variations as a new component~\citep{2005MNRAS.357..145S}, which
requires extra frequency channels. Alternatively, the frequency
channels can be degraded into many low-resolution, high
signal-to-noise pixels~\citep{2006ApJ...641..665E}, in which case an
approach for dealing with the high-$\ell$ modes is still required.

For the multi-pixel approach it is clear that the map could be
  divided up into many sub-regions in which the spectral index is
  estimated, thereby introducing extra foreground parameters at a cost
  of increased noise at low $\ell$.  Gaining a quantitative
  understanding of the impact of spectral index variations on the
  multi-pixel method is clearly required, since the constancy of the
  spectral parameters is a central assumption in deriving the spectral
  index likelihood Equation~(\ref{eqn:slopelikeMpix}), but is not the
  main focus of this work.

\section{Mock observations and foreground case studies}
\label{sec:mock}

We define here two fiducial experimental setups which we will use in
the following analysis. They are chosen to reflect some general
characteristics of bolometric experiments, but idealized and
simplified for demonstration purposes.  We emphasize that we do not
attempt here to forecast a performance of any specific experiment, but
rather, on one hand, to demonstrate the performance of the considered
methods in the context of the future B-mode experiments, and on the
other, to provide reference numbers quantifying the precision levels
potentially achievable by the small-scale experiments.

\subsection{Balloon-borne experiment}
We consider a balloon-borne experiment data set with the following
characteristics:
\begin{itemize}
\item {\bf Survey area:} Approximately $1\%$ of sky, 
corresponding to around $126,000$ $3.5'$ pixels, showed in 
Figure~\ref{fig:masks}. 
\item {\bf Frequency coverage:} Three channels at $150$, $250$, and
  $410$ GHz each with the same Gaussian beam of FWHM$= 8'$.
\item {\bf Noise level:} Homogeneous uncorrelated noise with an RMS of
1.5, 4 and 40 $\mu$K on a $3.5'$ pixel.
\end{itemize}
Such a {\bf setup} can be considered as a minimal choice for a
balloon-borne observatory. It takes an advantage of not being limited
in the frequency range by the atmosphere on one hand, and on the other
it restricts the number of dominant foreground components to one.  It
can be considered as an idealization of the data set anticipated
from, for instance, the EBEX experiment \citep{2004SPIE.5543..320O,
  2008SPIE.7020E..68G}.

We combine these experimental specifications with the following
three case studies:

\begin{description}
\item{\bf Basic: }{Based on the basic sky model and the observation characteristics
as defined above.}
\item{\bf Small-scale power:} The basic sky model is augmented with
  extra small-scale power in the dust.
\item{\bf Varying spectral index:} The basic sky model is
  augmented with a variation of the spectral index.
\end{description}

\subsection{Ground-based experiment}
Our ground-based experiment is characterized as follows:
\begin{itemize}
\item {\bf Survey area:} Approximately $2.5\%$ of the sky
  corresponding roughly to $320,000$  $3.5'$ pixels, showed in 
Figure~\ref{fig:masks}.
\item {\bf Frequency coverage:} Three channels at $90$, $150$, and $220$ GHz
  each with FWHM$= 8'$.
\item {\bf Noise level:} Homogeneous uncorrelated noise with an
RMS of 3, 3, and 9 $\mu$K on a $3.5'$ pixel.
\end{itemize}
The frequency range is chosen to fit in the window allowed by
the atmosphere on one hand, and on the other to conform with the
optimal working conditions for bolometric detectors. As a result, the
assumed, covered range is quite limited. We note that going beyond the
lower frequency bound can easily be imagined by combining radiometric
and bolometric data. We will however use the `minimal' setup defined
above as our standard case, and extend it on an as-needed basis.  We finally
note that such characteristics are not far
from those planned for, for instance, the first deployment of the 
POLARBeaR experiment \citep{2008AIPC.1040...66L}. 

The qualitative difference between the ground-based and balloon-borne
configurations is that the ground-based frequency channels are much
closer to our foreground minimum at approximately 70 GHz
where, by definition, two foregrounds are present. This leads us to consider
the ground-based experimental configurations with a more involved set of
foreground case studies:
\begin{description}
\item{\bf Basic:} Based on the basic sky model and the observation characteristics
as defined above.
\item{\bf Dust spectral index prior from balloon-borne experiment: }
Here we assume the value of the dust spectral index determined by the
balloon-borne experiment. We discuss two cases with all three or only
two ($150$ and $220$ GHz) channels included.
 \item{\bf Synchrotron template: } In this case
   each channel has been corrected for the presence of the synchrotron
   signal via subtraction of a external synchrotron template. We will
   assume that the subtraction is performed down to some predefined
   precision level. 
\item{\bf Extra low frequency channel: } Here we add an extra channel
  centered at $40$ GHz to the basic data set with the noise as in the
  $90$GHz map case.
 \end{description}
We point out that only one case above (the basic case) is both
realistic and self-contained. In all the other cases, the
presence of some extra external knowledge is postulated.

 In addition to the cases listed above we have also performed the same
 analysis assuming a lower noise level of $1 \mu$K per $3.5'$ pixel in 
all three channels, as well as devised specific test runs designed to 
 highlight some particular aspects of the performance of the component
 separation method.  These include:
 \begin{description}
\item{\bf No synchrotron:} In this case the problematic
synchrotron emission is removed from the sky model in order to
understand the way in which it biases the dust estimation and
subtraction.
\item{\bf Shuffled synchrotron template:} In this case the
  spatial morphology of the synchrotron template is randomized. The
  idea here is to understand the effect of accidental correlations
  between the foreground components in this regime of a restricted
  number and coverage of the available frequency channels.
\end{description}

\begin{figure}
\begin{center}
  \includegraphics[width=7.5cm]{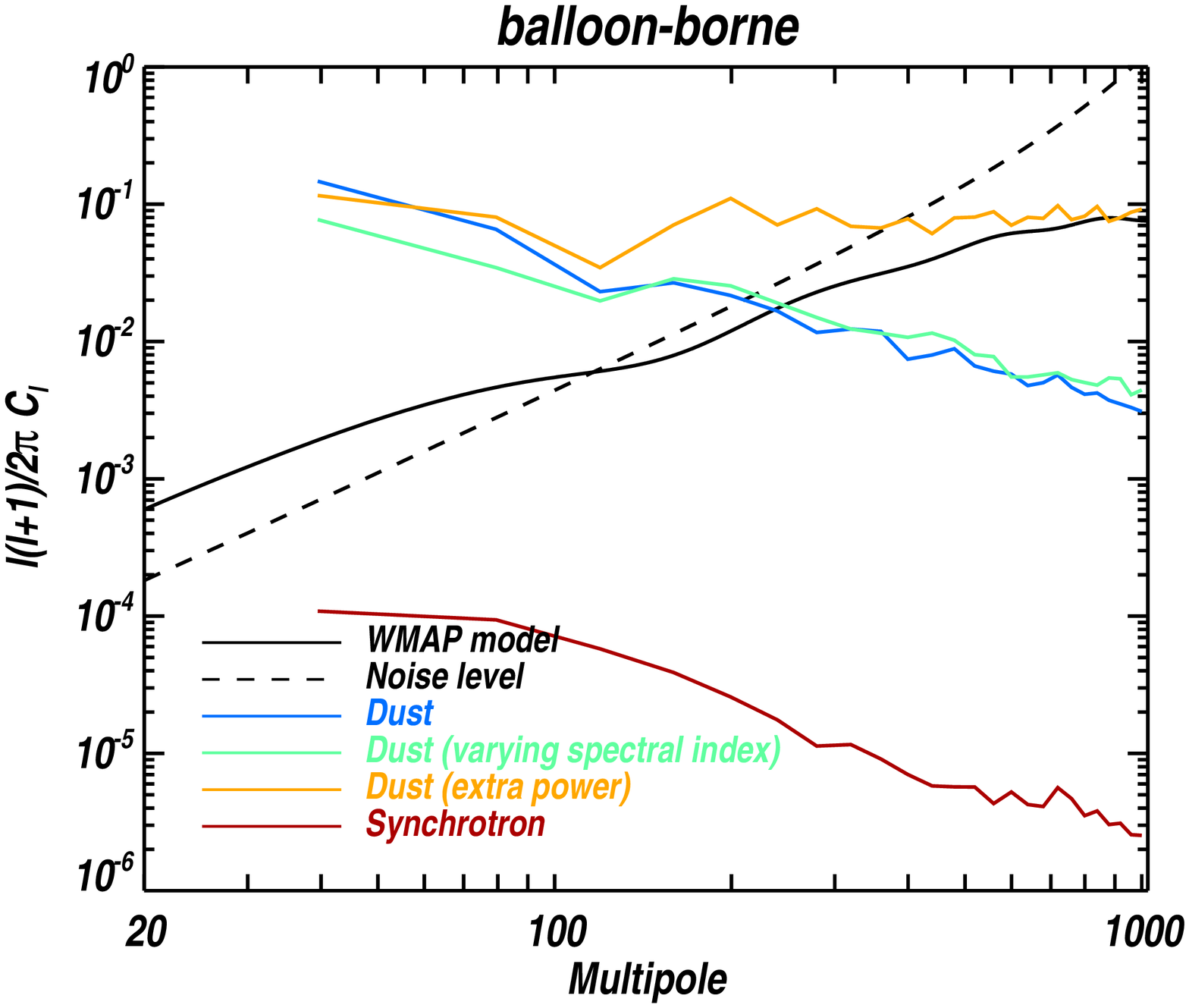}
  \includegraphics[width=7.5cm]{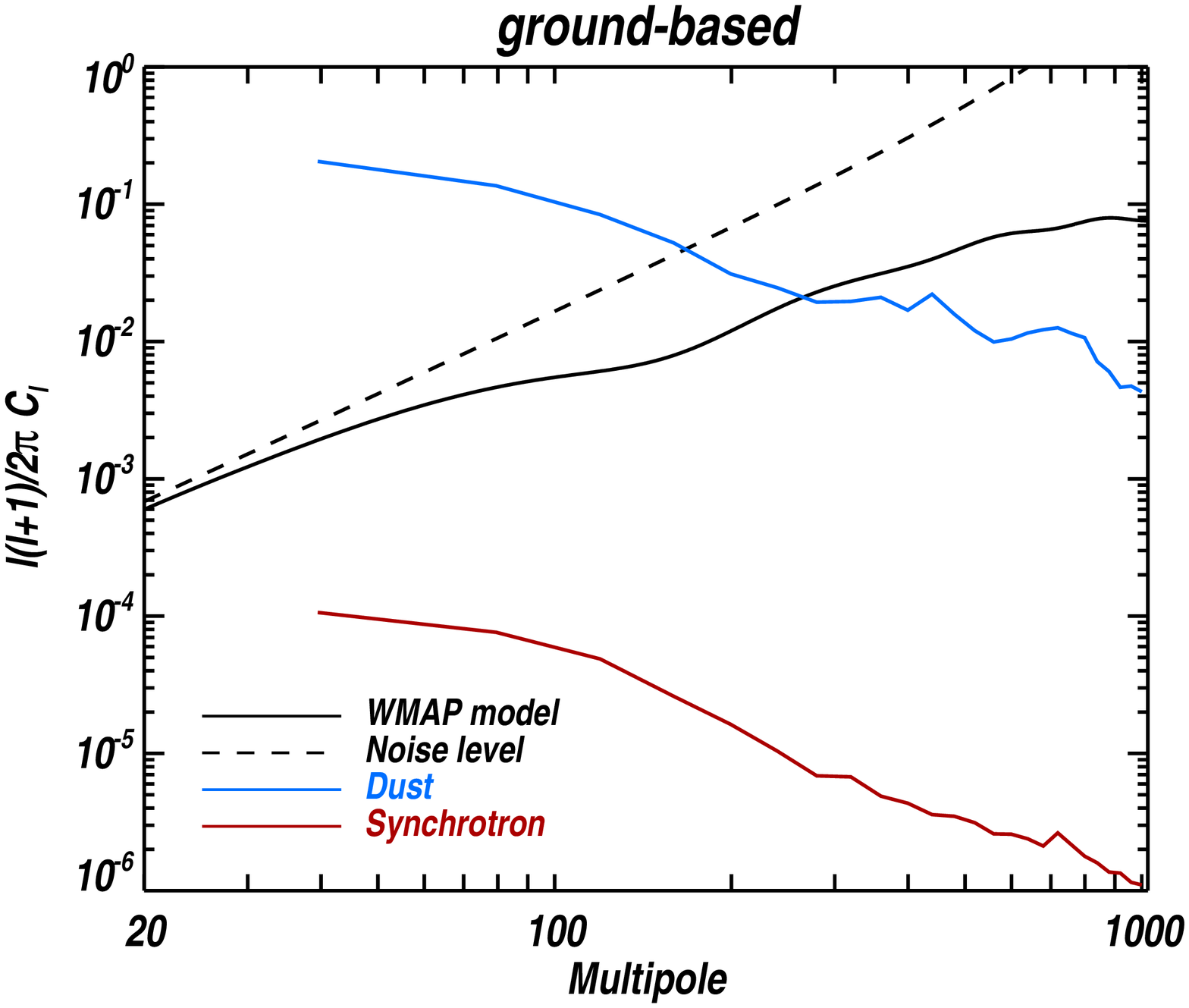}
  \caption{B-mode spectra of the input components at the 150 GHz
           channel, for the balloon-borne (upper) and ground-based (lower
      panel) experiments. }
    \label{fig:input_bmodes}
\end{center}
\end{figure}

We show in Figure~\ref{fig:input_bmodes} the B-mode power spectra of
the input components at 150 GHz. Depending on angular scale, the dust 
amplitude must be
suppressed by between a factor of 5 and 25 to have successful
measurement of the cosmological B-mode signal.

  We note again that the noise levels of the derived CMB maps in the
  basic balloon-borne and ground-based cases assuming a perfect
  knowledge of the foreground frequency scaling properties are 1.6
  and 3.2 $\mu$K per $3.5'$ pixel, respectively (while in the low noise 
  ground-case we get 0.9 $\mu$K).
  These can be obtained
  from Equation~(\ref{eqn:noiseCorrOptim}) and are used for
  the noise level shown in Figure~\ref{fig:input_bmodes}.  

\subsection{Foreground spectral modelling}
\label{spec_model}

Implicit in our approach is the assumption that the three channel
configurations of the ground-based and balloon-borne just described
will provide information on three, but no more than three, parameters:
the CMB amplitude and the dust amplitude at each pixel, and the dust
spectral index, as constrained by the ensemble of data and by two
Stokes parameters. In deriving results in the following section, we
begin by assuming exactly the same smooth model,
Equation~(\ref{eq:dust_scal}), for fitting the dust as was used in the
simulations. To some degree this choice is made for expediency, in
order to define a comparison benchmarks for our case studies, and in
order to disentangle the effect of different factors on the quality of
the CMB estimation. We do however attempt to gauge the strength of
this assumption by studying two further case studies that are relevant
in this context. These are map-level calibration uncertainties and
`spectral mismatch'. Intuitively, calibration uncertainties will
downgrade our ability to infer useful information about the dust model
and spectral index. Spectral mismatch refers to multiplicative factors
applied to the dust model at each channel, breaking the smoothness of
the dust emission spectra. This situation could physically occur
  when the dust scaling can not be sufficiently well characterized by
  the three channel experiment. For instance one can conceive of a
  case in which two dust components with warmer and colder
  temperatures also have different polarization fractions, leading to
  a sudden change in the dust scaling (and/or position angle)
  coincident with the frequency coverage of the experiment. However, this
  particular case is thought to be not particularly likely for
  nearby Galactic cirrus, as reviewed by \citet{2009AIPC.1141..222D}.
  Spectral mismatch could also result from poorly characterized
  experimental bandpasses~\citep{2003ApJ...582L..63C}.

\section{Results}
\label{sec:Results}

\subsection{Performance evaluation metric}

To assess the performance of our component separation method we will
look both at the estimated foreground spectral indices and at the quality of the
recovered CMB maps. 
The latter are clearly not expected to be perfect with potential contamination
arising either due to the noise present already in
the input, single channel maps, or a failure of the algorithm to
perform the separation perfectly. This may result either in some
level of non-CMB signal still present in the map, or in the CMB signal
being compromised.  These two effects are usually referred to as
residuals. With the noise uncertainty being quite straightforward to
characterize using Equation~(\ref{eqn:noiseCorrOptim}), it is our aim to
evaluate the level of the residuals expected in the foreground case studies
and then to compare it with statistical uncertainties.  The
latter includes just pixel noise in the case of the maps, and both the
noise and CMB signal variance on the level of the power spectra
analysis.

For each of the case studies described in the previous section
we first estimate the best-fit spectral parameters by maximizing the
spectral likelihood, Equation~(\ref{eqn:slopelikeMpix}), including on
occasions some extra prior information. Then, given the estimated
values of the spectral parameters we compute the map of residuals as,
\begin{eqnarray}
\label{residuals}
\bd{\Delta} = \bd{\hat s}-\bd{s_0}- \l(\bd{\hat{A}}^t\,\bd{N}^{-1}\,\bd{\hat{A}}\r)^{-1}\,\bd{\hat{A}}^t\,\bd{N}^{-1}\,\bd{n},
\label{eqn:resDef}
\end{eqnarray}
where $\bd{s_0}$ are the input simulated components and the last
term on the right hand side subtracts away the noise in the recovered
components $\bd{\hat s}$, given the best-fit values for the spectral
parameter and the known input noise realization $n$. Directly
subtracting the input noise improves upon the metric used by
\cite{2008A&A...491..597L} in which residual noise was suppressed by
smoothing. The CMB element of Equation~(\ref{residuals}), $\bd{\Delta}^{\rm
  CMB}$, quantifies the residual foreground signal contained in the
estimated CMB map, as well as part of the genuine CMB signal
correctly interpreted by the algorithm as the CMB contribution.
Finally, we compare the latter with the anticipated level of the
genuine CMB B-mode signal as well as level of its statistical
uncertainty due to only CMB and noise sampling and cosmic variance.
For the last step we use as a metric the B-mode power spectra
calculated with help of the pure estimator. As described in Section~\ref{sec:StatVariance}, the spectrum variance is
estimated via $500$ Monte-Carlo simulations, for which
we use the best-fit WMAP 5-year cosmology with $r=0.05$ as the fiducial
model. A satisfactory level of foreground cleaning is achieved
  when the foreground residual power spectrum is smaller than the
  statistical uncertainties, ensuring that the systematic errors due
  to foreground contamination are subdominant to the global error
  budget. 

In Section~\ref{sect:anaRes} we perform an analysis of these
   residuals, and in Section~\ref{sec:tensor2scalar} we express the
   results in terms of an {\em effective} detectable value of the
   tensor-to-scalar ratio, $r$.

\subsection{Balloon-borne cases}

In Table~\ref{beta_balloon} we report the recovered values of the 
dust spectral index $\beta$ for the three tests we have made: in all
the cases this parameter was successfully estimated.

\begin{table}
\begin{center}
\caption{Results for the dust spectral index estimation in the
balloon-borne case. The last column give the RMS of the total
foreground contamination left in the CMB map. The input dust spectral
index is $\beta_d=1.65$. For comparison, the B-mode signal RMS on the 
pixel scale is 0.19$\mu$K.}
\label{beta_balloon}
\begin{tabular}{cccc}
\hline
\multicolumn{4}{c}{\bf Balloon-borne ($\beta_0=1.65$)}\\
\hline
\ Case & $\beta$ & $\Delta \beta$ & RMS ($\mu$K)  \\
\hline
\ Basic & 1.655   & 0.009 & 0.020\\
\ Small-scale power & 1.655 & 0.009& 0.021\\
\ Varying spectral index & 1.657 & 0.011& 0.024\\
\hline
\end{tabular}
\end{center}
\end{table}

\begin{figure*}
\centering
\includegraphics[width=6.5cm]{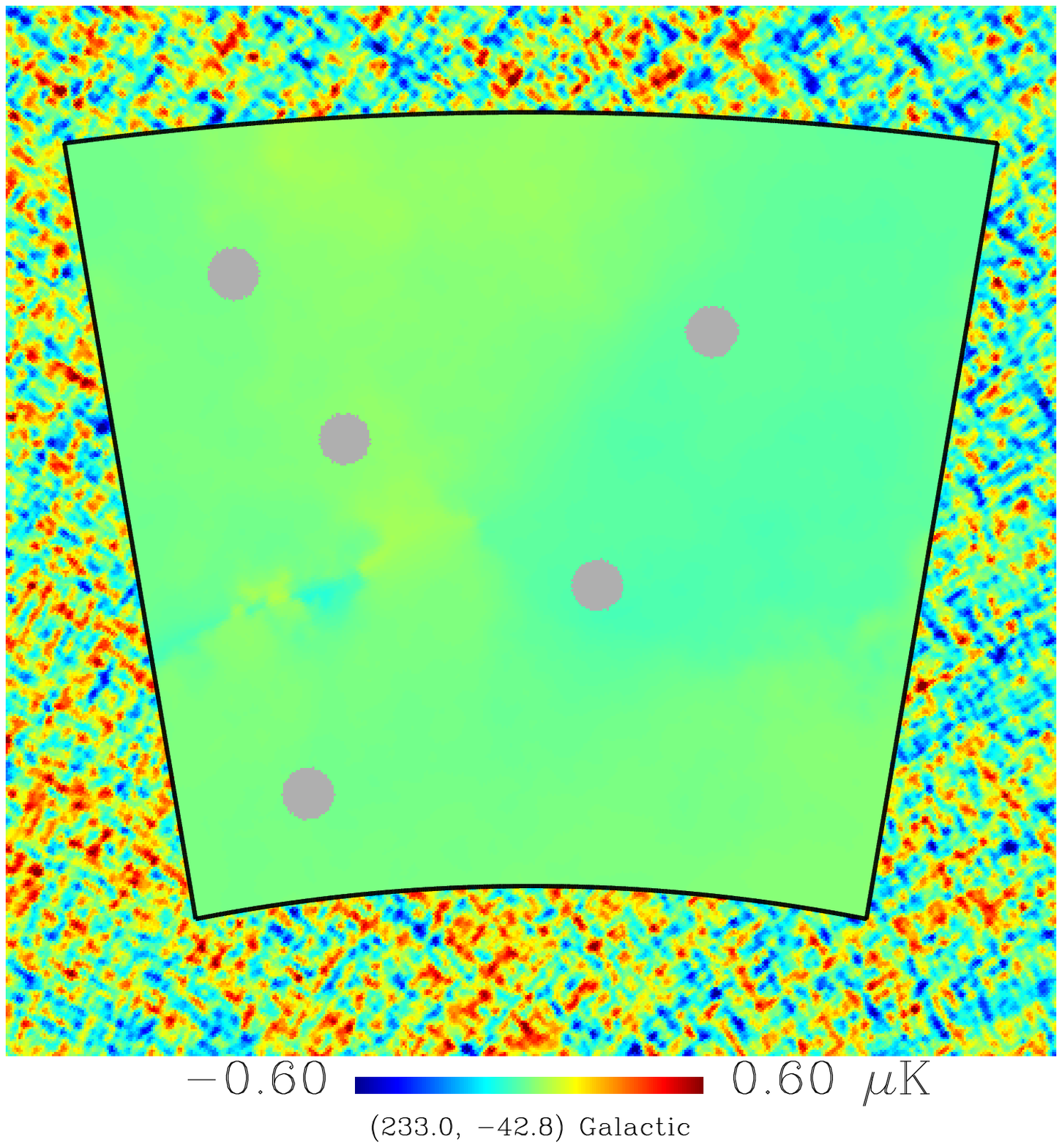}
\includegraphics[width=6.5cm]{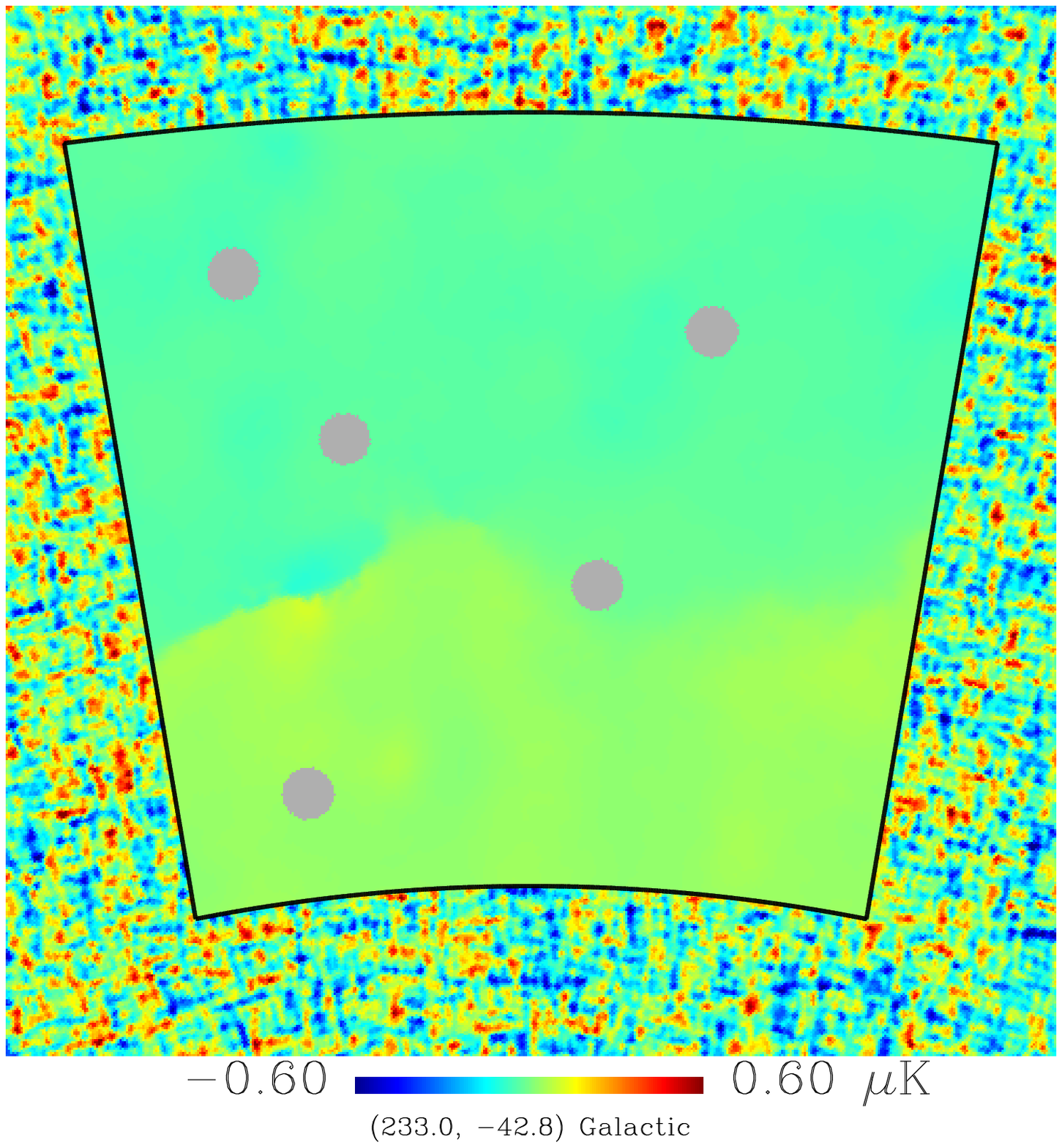}
\caption{{\bf Balloon-borne experiment.} Maps of residuals in the CMB map, as
  defined in Equation~(\ref{eqn:resDef}), for the Stokes $Q$ (left)
  and $U$ (right) parameters in the basic case. For comparison, outside the survey
  border is shown a pure B-mode realization.}
\label{Cases_balloon}
\end{figure*}

In Figure~\ref{Cases_balloon} we show the residuals, $\bd{\Delta}$, for
the basic case. To appreciate the level reached by the cleaning, they 
can be compared
with the input foreground contamination at 150 GHz, shown in
Figure~\ref{sky_template}.  The resulting residuals for all the
balloon-borne cases differ only in a minor way and they are always 
dominated by the
unmodelled synchrotron because the accurate estimates of
the dust spectral index, as derived earlier,  allow the dust to be
subtracted with a superior precision.

\subsection{Ground-based cases}
\label{subsect:groundBasedCases}

To perform the foreground cleaning, we again assume the presence of a
single dust foreground because the limited number of channels prevents
us from performing spectral modelling of the synchrotron present
in the data.

\begin{figure*}
\centering
\includegraphics[width=6.5cm]{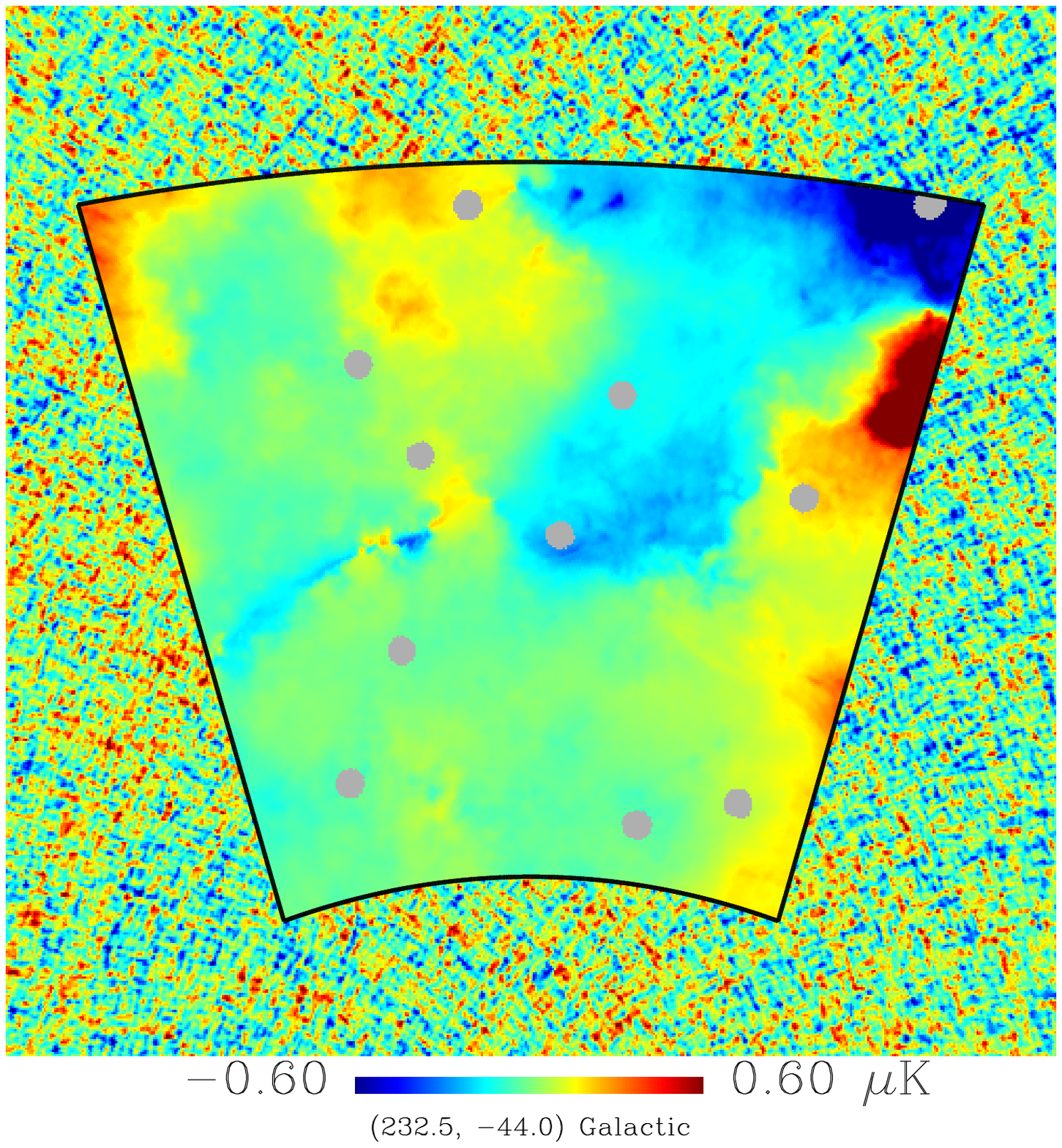}
\includegraphics[width=6.5cm]{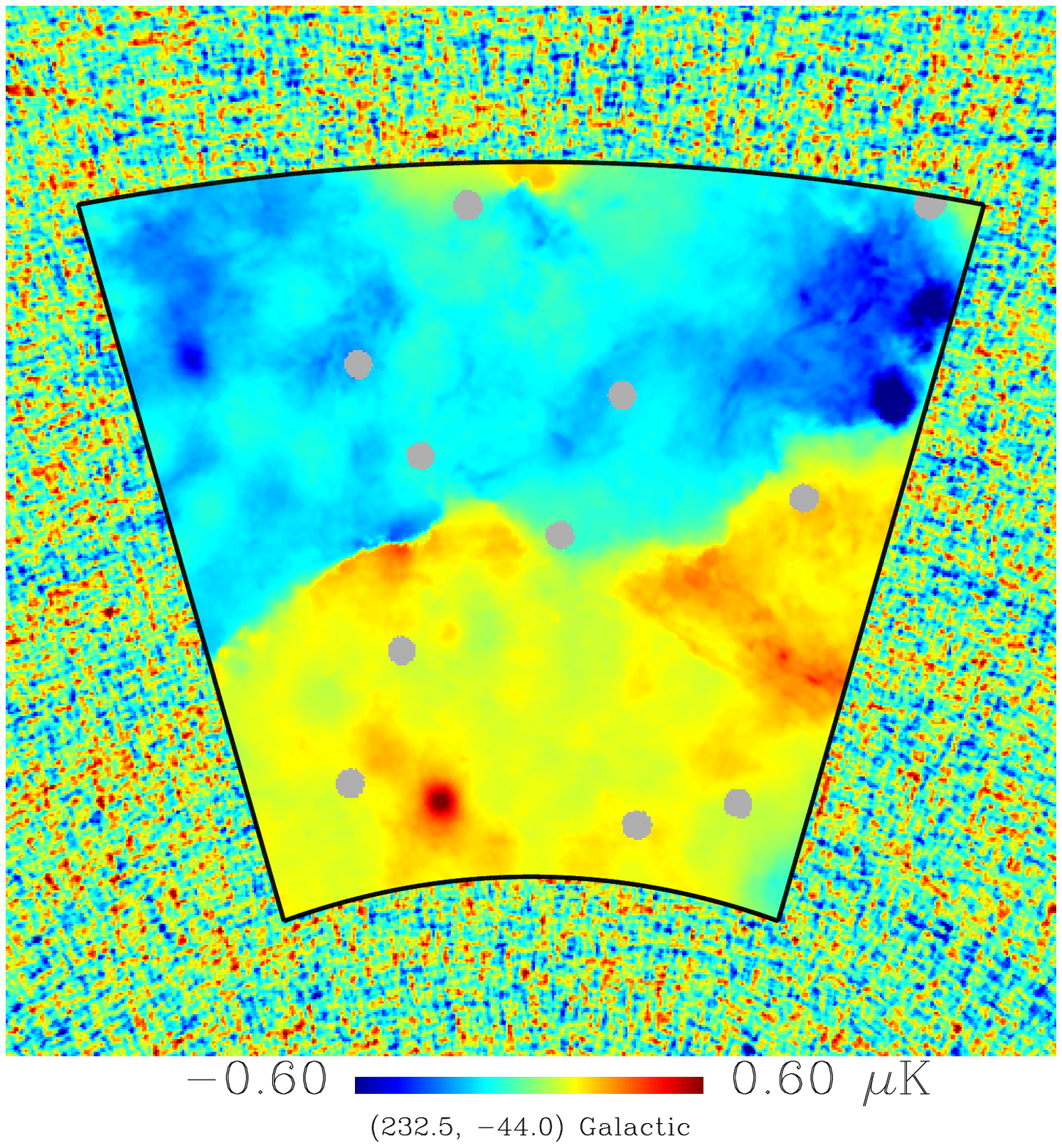}
\caption{{\bf Ground-based experiment.} Maps of residuals in the CMB map for the
  Stokes $Q$ (left) and $U$ (right) parameters in the basic case, as in Figure~\ref{Cases_balloon}.}
\label{Cases_ground}
\end{figure*}

The basic result, reported in Table~\ref{beta_ground}, is that
the estimated dust spectral index $\beta= 1.875 \pm 0.028$ is
significantly biased away from the input value of 1.65, 
giving rise to residuals significantly higher 
than found in the balloon-borne cases, as shown in Figure~\ref{Cases_ground}
and quantified by an RMS larger by a factor $\sim 8$.
This bias of the dust
spectral index can be explained qualitatively by the fact that in this case
not only does the synchrotron component, which remains unmodelled and unsubtracted, have a higher
amplitude than before due to a presence of the 90 GHz channel, but also because it has a significant spatial
correlation with the dust component, as shown in
Figure~\ref{sky_template} and discussed in Section~\ref{sec:simulated_sky}. 
We have verified this explanation by making
tests first with no synchrotron present and later including only
the `shuffled synchrotron'. The latter case erases the
synchrotron-dust correlation, artificially converting the synchrotron
to a white noise-like component with less fluctuations on large
angular scales. In both these artificial test cases, satisfactory
spectral index estimates and
foreground cleaning were obtained (see second and third panel of 
Figure~\ref{Cases_ps_ground}). Moreover we found that 
the pixel by pixel correlation between the dust and synchrotron, computed with the Pearson coefficient 
$C={\rm cov}\l(X,Y\r)/\sigma_x\sigma_y$, has to drop below $\sim 15\%$
to allow for satisfactory foreground cleaning.

\begin{table}
\begin{center}
\caption{Results for the dust spectral index estimation in the
  ground-based case, as in Table~\ref{beta_balloon}. The
  asterisk indicates the cases in which the dust spectral index is
  fixed, based on the value recovered from the balloon-borne
  experiment.}
\begin{tabular}{cccc}
\hline
\multicolumn{4}{c}{\bf Ground-based  ($\beta_0=1.65$)}\\
\hline
\ Case & $\beta$ & $\Delta \beta$ & RMS ($\mu$K) \\
\hline
\ Basic & 1.875 & 0.028& 0.170\\
\ No synchrotron & 1.642 & 0.030 & 0.004 \\
\ Shuffled synchrotron & 1.667 & 0.028 & 0.074 \\
\hline
\ $90$+$150$+$220\;$+ balloon expt. & $1.655^*$ & $ - $ & 0.077\\
\ $150$+$220\;$+ balloon expt.  &  $ 1.655^* $ & $-$ & 0.034\\
\ External template   & 1.682 &  0.026 & 0.028\\
\hline
\end{tabular}
\label{beta_ground}
\end{center}
\end{table}
 
\begin{figure*}
\centering
\includegraphics[width=5.8cm]{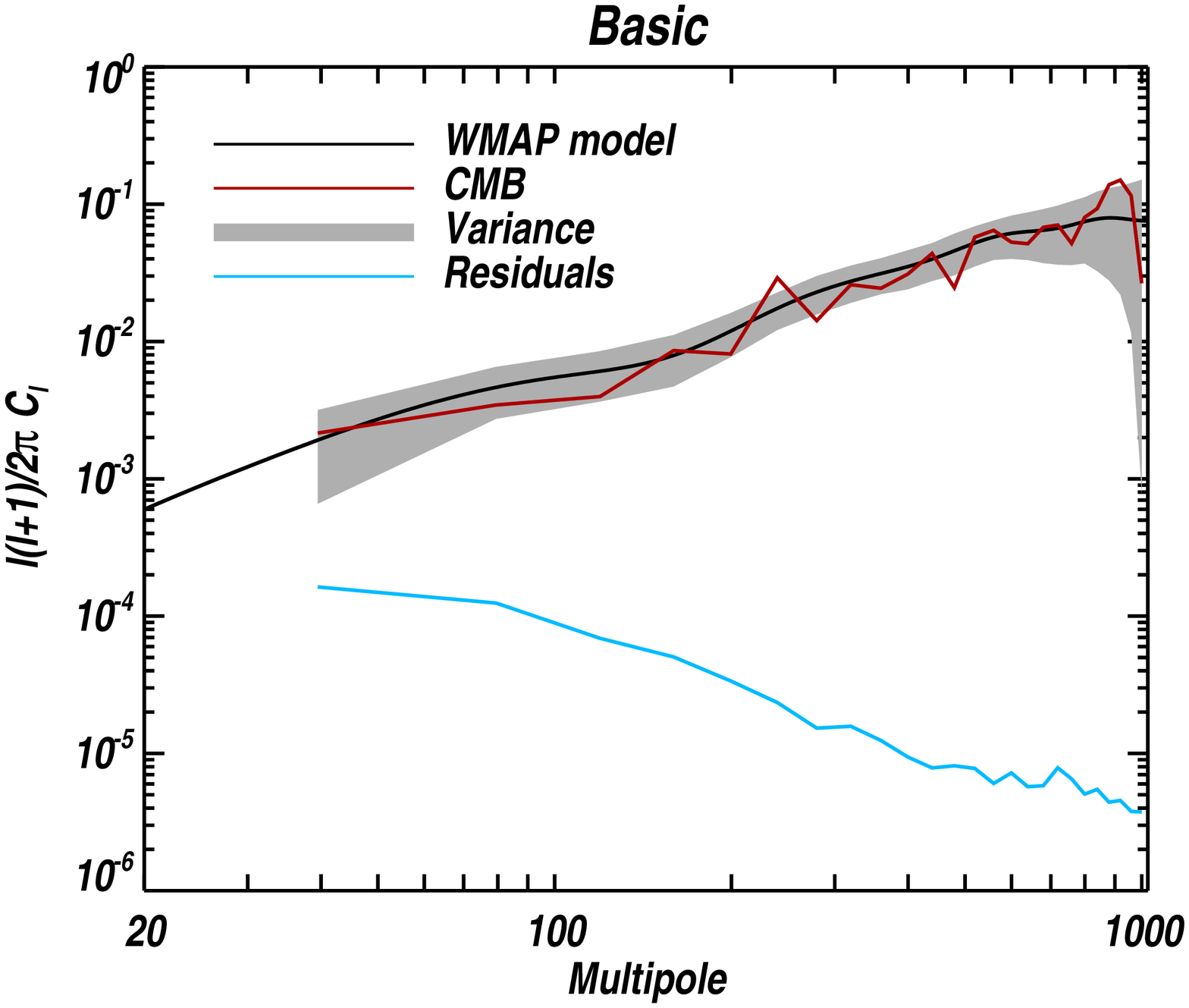}
\includegraphics[width=5.8cm]{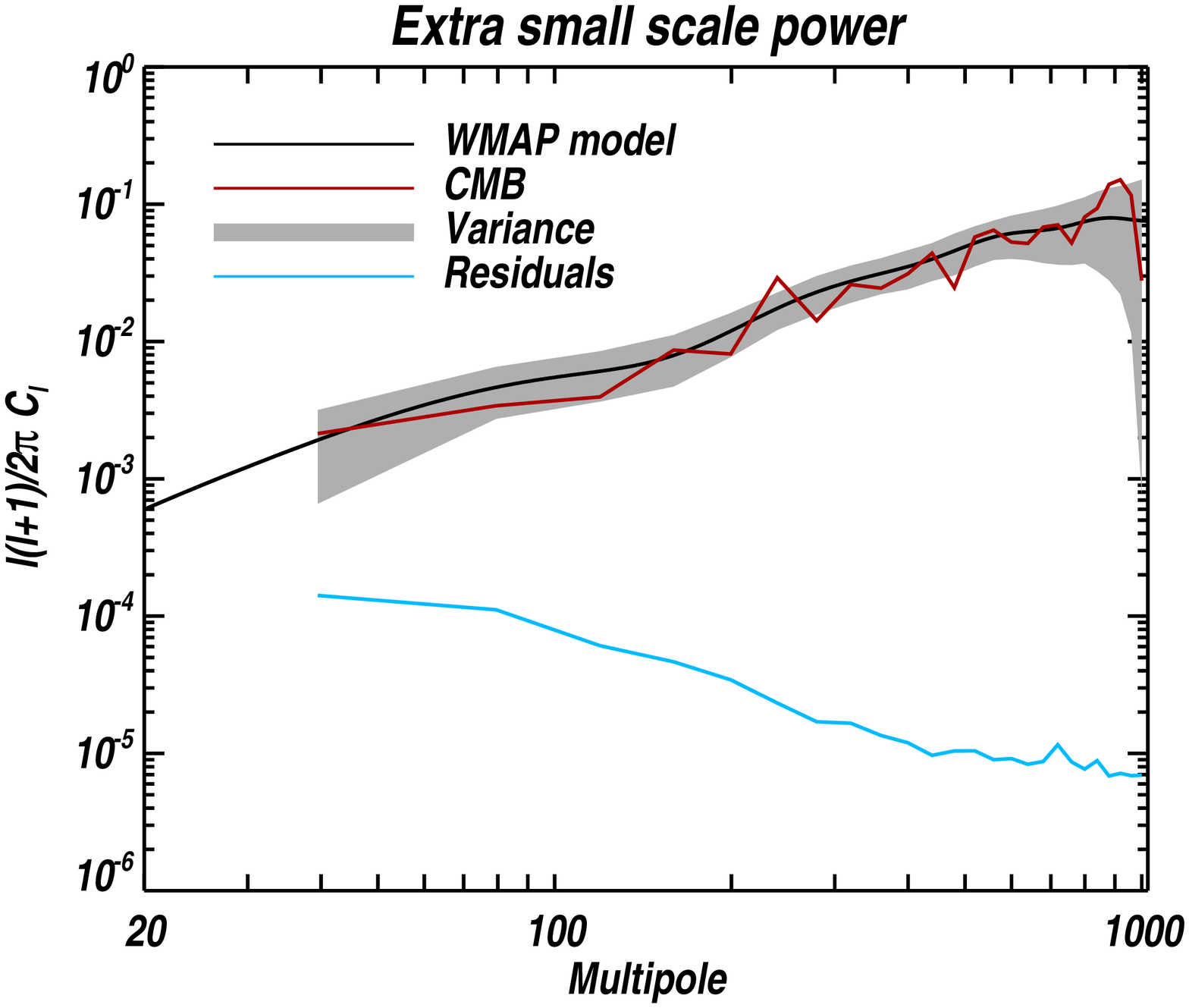}
\includegraphics[width=5.8cm]{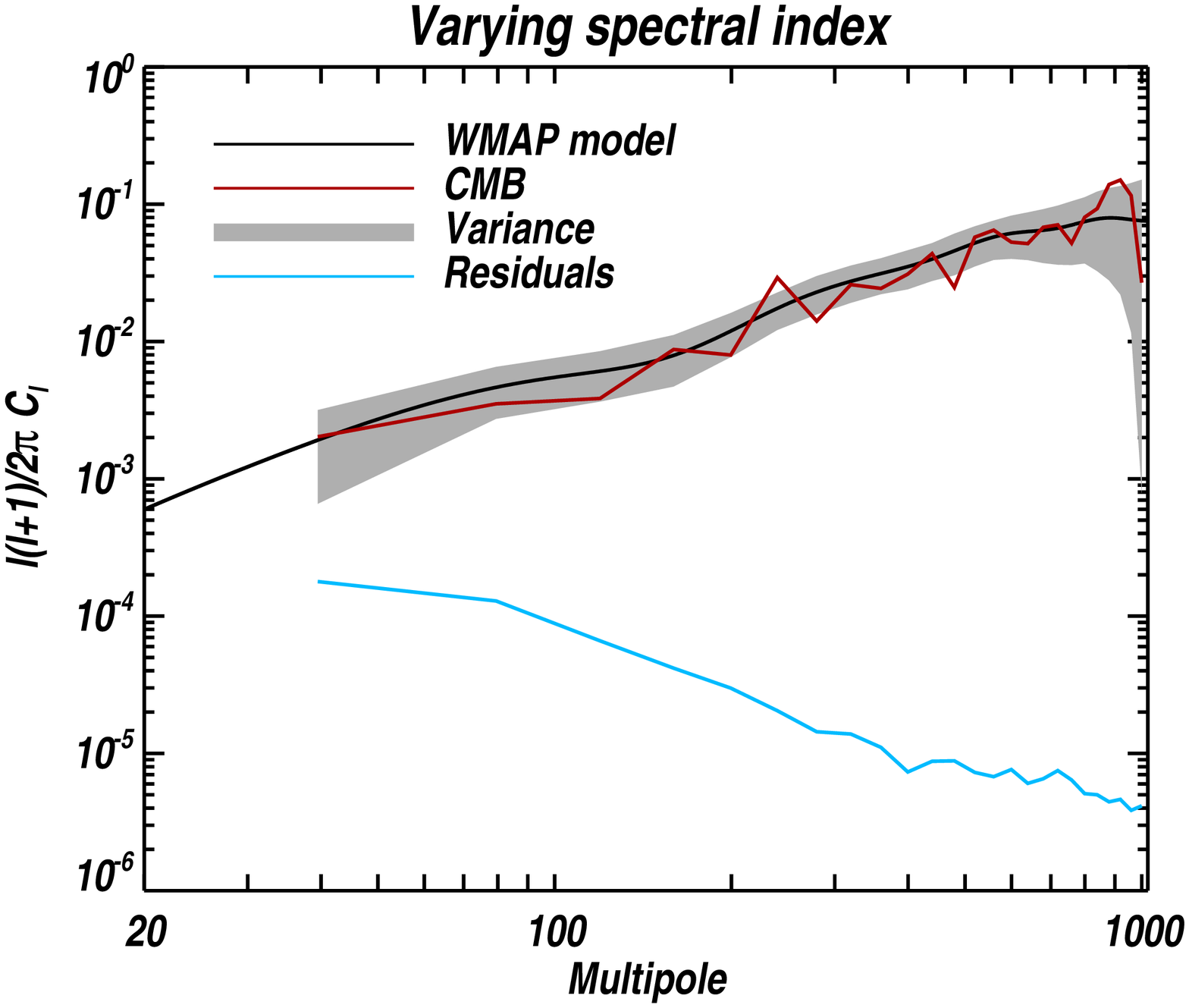}
\caption{{\bf Balloon-borne experiment.} B-mode
  power spectra of the residuals in the CMB map (cyan curves). As a reference, the input and estimated
  CMB B-mode power spectra are represented by solid black and the solid
  red curves. The statistical uncertainty of the CMB power
  spectrum estimation is shown by the grey band. For both
  E- and B-modes and for all these cases, the level of foreground
  residuals is smaller than the statistical uncertainties ensuring a
  precise enough foreground cleaning. From left to right: i)
  The basic case, ii) Extra small-scale power for the dust emission, 
iii) Spatially-varying spectral index for the dust emission. }
\label{Cases_ps_balloon}
\end{figure*}

\begin{figure*}
\centering
\includegraphics[width=5.8cm]{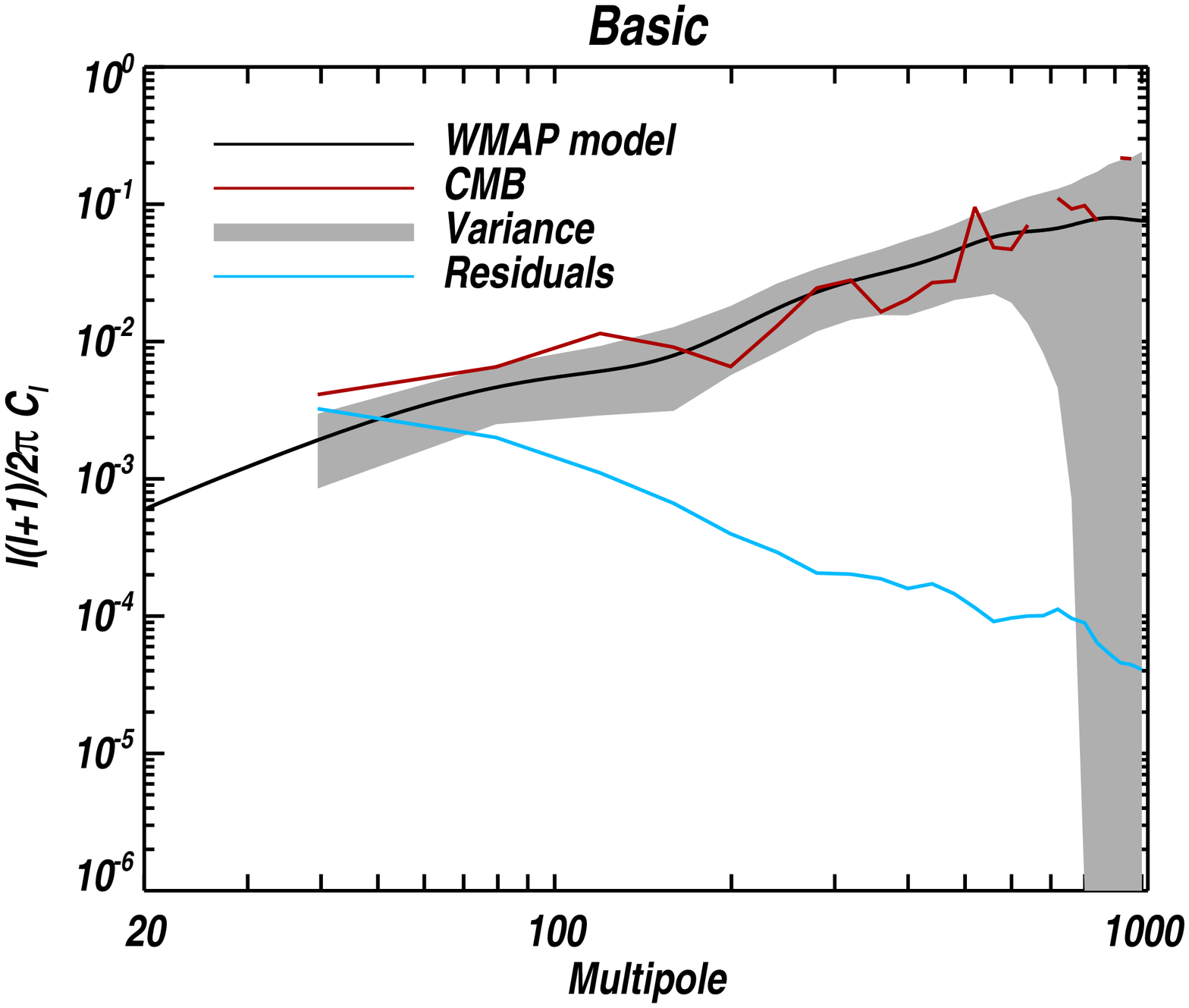}
\includegraphics[width=5.8cm]{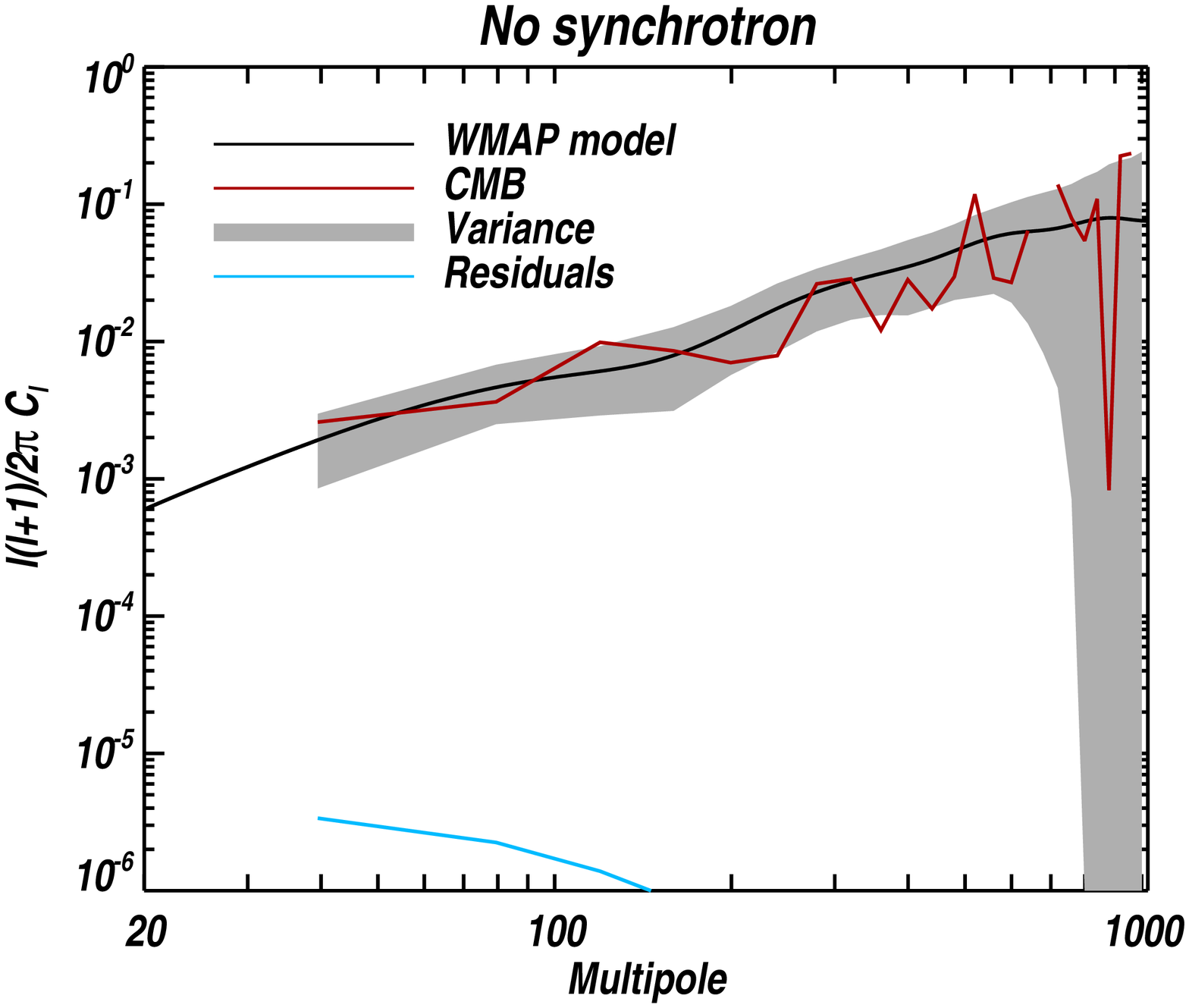}
\includegraphics[width=5.8cm]{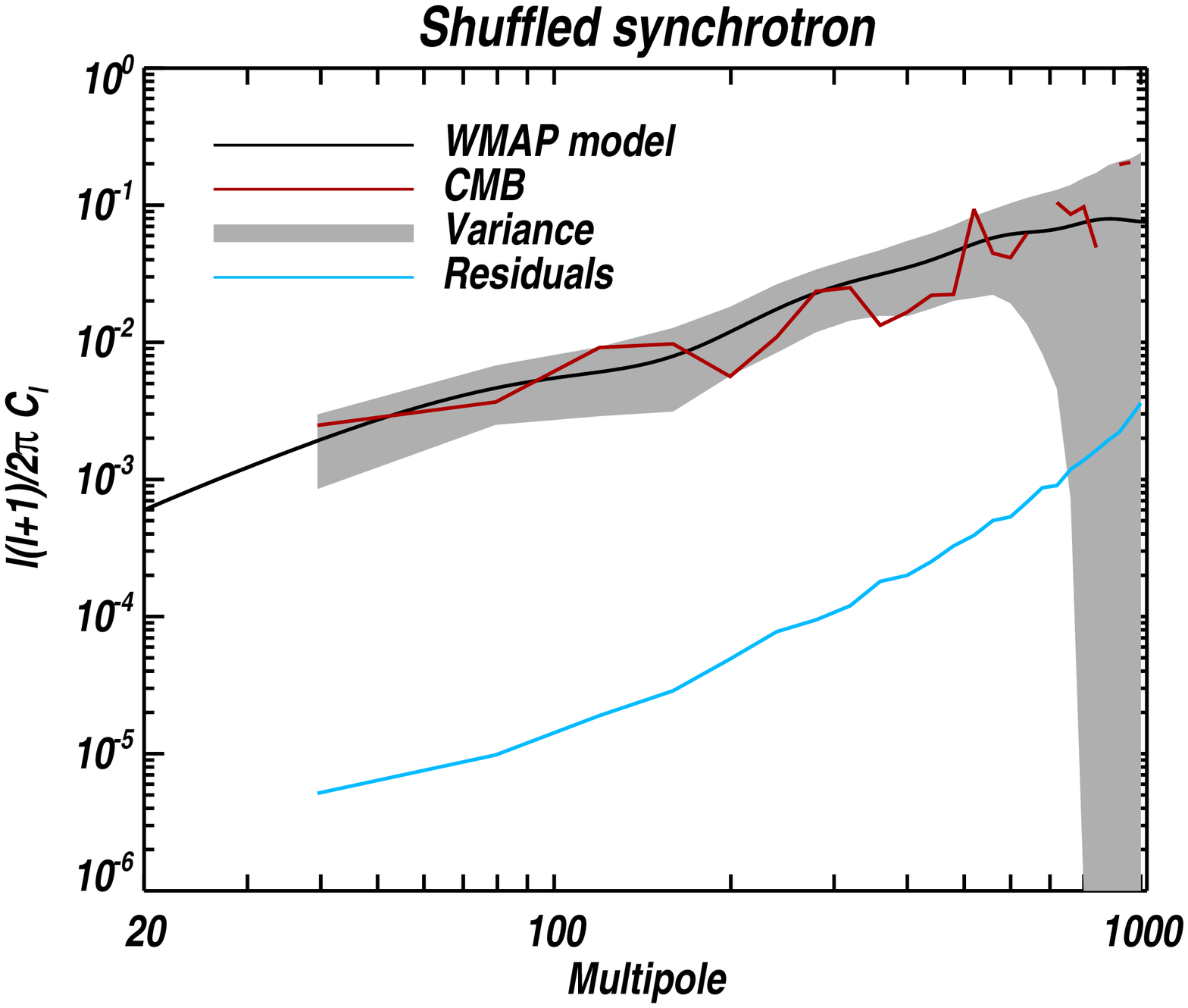}
\caption{{\bf Ground-based experiment.} B-mode power spectra of
  the residuals in the CMB map, as in
  Figure~\ref{Cases_ps_balloon}. From left to right: i) The basic case,
  ii) A test case with no synchrotron in the simulation, iii) A test
  case in which the pixels in the synchrotron map were reshuffled to 
remove the spatial
  correlation with the dust.}
\label{Cases_ps_ground}
\end{figure*}

\subsection{Ground-based cases with external information}
The ground-based setup discussed here is therefore not
self-sufficient and thus unable to provide a appropriately 
cleaned CMB map. In
this Section we therefore investigate the effect of using `external
information', specifically priors on the foreground spectral indices
or an external synchrotron template, on the analysis of 
this data set.

\begin{figure*}
\centering
\includegraphics[width=5.8cm]{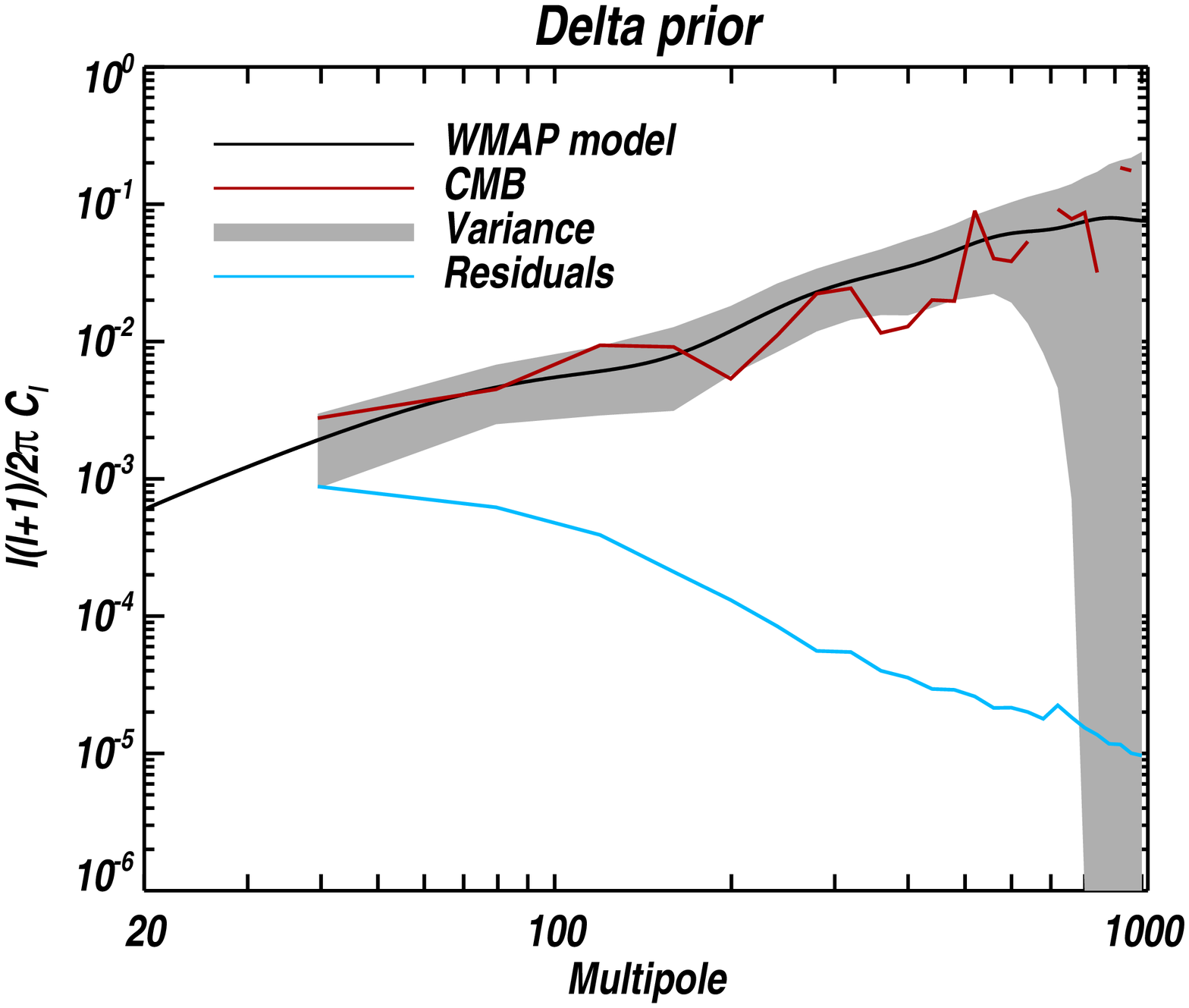}
\includegraphics[width=5.8cm]{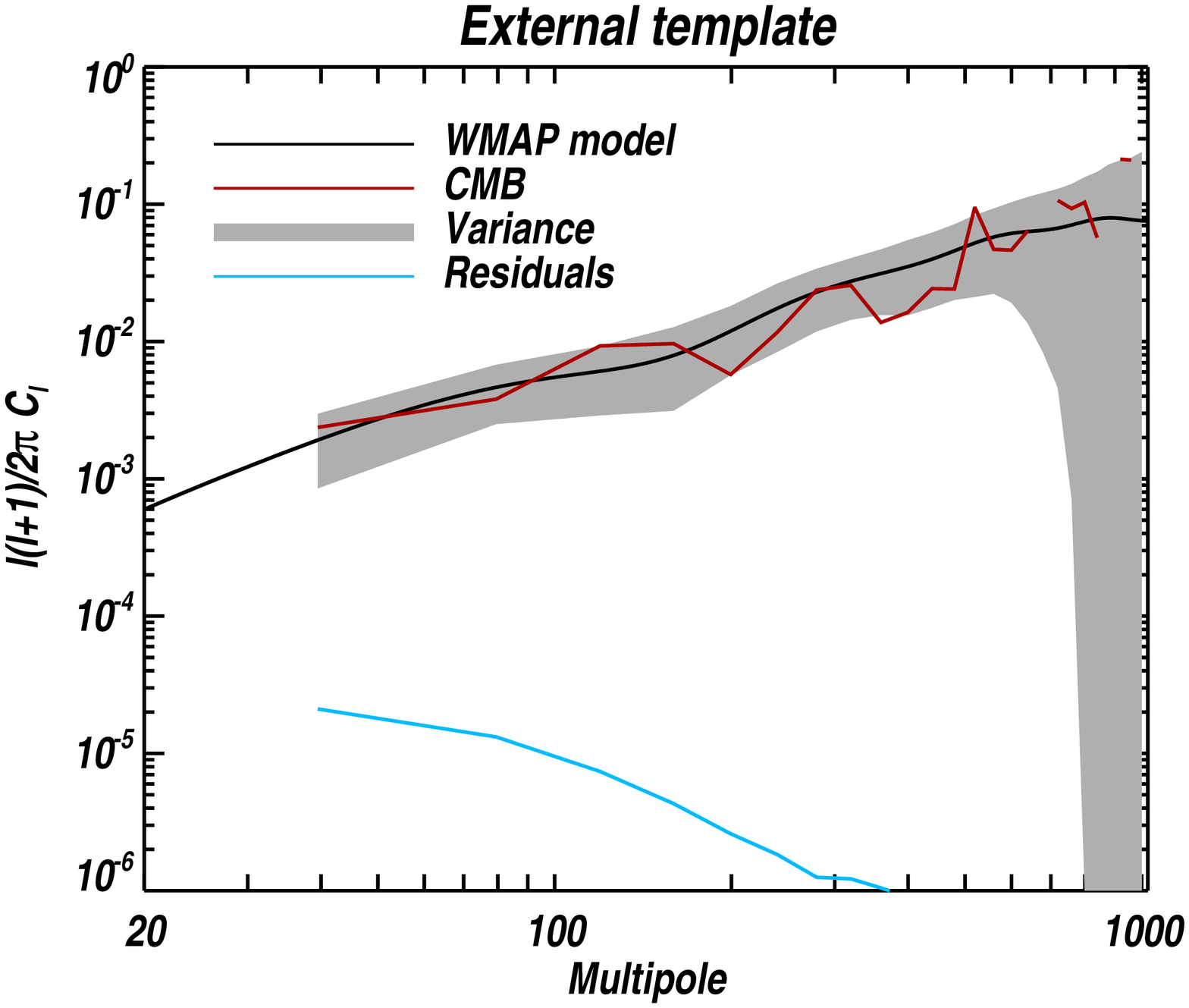}
\caption{{\bf Ground-based experiment.} B-mode power spectra of the residuals in the CMB map, as in
  Figure~\ref{Cases_ps_balloon}. Left: The basic case, imposing the
  the dust spectral index value recovered by the balloon-borne
  experiment (to be compared with the first panel of Figure~\ref{Cases_ps_ground}). Right: Subtracting a synchrotron template with an amplitude known 
to within a $10\%$.}
\label{Cases_ground_ps_cure}
\end{figure*}

The first test we made, mimicking a possible real life situation, was
to impose strictly a dust spectral index prior with the value found in the
balloon-borne basic case ($\beta=1.655$). At this point, there are no
free spectral parameters to estimate, and the corresponding least
squares components can be  directly estimated. The residuals for this case are
shown in the `delta prior' panels of Figure~\ref{Cases_ground_ps_cure}, which
demonstrates that the dust spectral index prior does help
reduce to some extent the residuals and the final spectral
contamination of the B-mode spectrum.  Thanks to the high precision 
of the estimation of the dust spectral index in the balloon-borne
experiment, we find that those residuals are again due to the unmodelled synchrotron
(on which we will elaborate in Section~\ref{sect:anaRes}).

Knowing that the 90 GHz channel is contaminated by synchrotron,
we also have investigated the impact of simply dropping this channel, fixing
again the dust spectral index to the value determined by the
balloon-borne experiment. Though clearly rather drastic, this choice could 
in principle
provide a better foreground cleaning than the three
channel setup. Unfortunately for the specific case analyzed in this work, 
we found that the remaining two channels are too noisy to produce a
 CMB cleaned map
suitable for any B-mode work. We also however have found that two-channel 
setup could
be a viable option if the noise in these two channels is suppressed down 
to a $\sim 1\mu$K level --
a rather challenging goal.
Nevertheless this result suggests that some
specific attention  may need be paid to finding the best trade-off between 
frequency bands
choice and observation time for this kind of experiment.

We also attacked the problem from the other side of the
foreground minimum, using information coming from lower
frequencies. First, we made use of an external template for the
synchrotron, whose amplitude is assumed to be known with a $10\%$
uncertainty, subtracting it directly from the data channels. The
satisfactorily foreground cleaned result for this case, in which no
prior on the dust spectral index was assumed, is shown in the second
panel of Figure~\ref{Cases_ground_ps_cure}.
We note that though in this test we have assumed a high resolution template
as only the low-$\ell$ modes need to be corrected, a low resolution
synchrotron templates, as for example anticipated from the CBASS
experiment\footnote{\tt www.astro.caltech.edu/cbass/C-BASS\_official\_site/Home.html}
should be sufficient. Also, in cases where the overall calibration of
the available template is considered less reliable than its morphology 
the template marginalization could be a more robust technique to be used 
in this context \citep{2004ApJ...615...55J}.

Yet another option we have considered is to extend  the covered frequency range
by adding an extra frequency channel operating at $40$ GHz. This could be achieved for 
example by co-analyzing the data of two bolometric and radiometric experiments observing
an overlapping sky area. In our analysis the extra $40$ GHz was assumed to have a similar 
noise level as the one at $90$ GHz. We have indeed found that such a combined 
analysis fares well in terms of residuals amplitude, which are comparable 
to those shown in the second panel of Figure~\ref{Cases_ground_ps_cure}, in spite
of the fact the found best-fit value of the spectral index, $\beta\simeq 1.75$, is still found to be 
significantly away of the true input value. This fact is illustrated in Figure~\ref{wa_beta} and discussed
in Section~\ref{sect:anaRes}. We note also that the gain from
the extra channel in the total noise budget of the final CMB map,
Equation~(\ref{eqn:noiseCorrOptim}),  turns out to be
negligible. 

To recap the results of this section so far, the balloon-borne
experiment represents an example of a self-contained case where, owing
to the fact that its channels are slightly displaced from the
foreground minimum, a single simplistic foreground can be adequately
characterized and subtracted. It also helps the ground-based
experiment to alleviate the effect of biases caused by spatial
correlation between the foregrounds. However it is easy to envisage
the need of the ground-based experiment for further external
information on the synchrotron in the form of a template or a
lower frequency channel.

\subsection{Miscalibration and spectral mismatch}
\label{sec:spec_mismatch}

\begin{table}
\begin{center}
\caption{This table summarises how miscalibration errors of the
    two higher frequency channels relative to the lowest frequency channel
  increases the foreground residuals $\Delta$ with respect to the basic case.}
\label{tab:miscalib}
\begin{tabular}{cc}
\hline
\smallskip
\ Channel miscalibration & ${\displaystyle \frac{\Delta_{\rm calibration}}{\Delta_{\rm basic}}}$  \\
\hline
\ $0\%, 2\%, 2\%$ &  $\sim 2$  \\
\hline
\ $0\%, 5\%, 5\%$ &  $\sim 3$  \\
\hline
\ $0\%, 10\%, 10\%$ & $\sim 4 $  \\
\hline
\end{tabular}
\end{center}
\end{table}

Here we extend the study conducted so far by incorporating two
  specific systematic effects which are miscalibration of the data
  channels and spectral mismatch between the assumed and the real
  model of foregrounds.  These two effects differ not only as to their origin, one
  being due the instrument properties and the other reflecting
  our ignorance of the physical phenomena relevant to the 
    following case studies.  They also appear differently
  within the discussed component separation formalism, within which a
  consistent description of only the miscalibration can be
  incorporated and thus its effects properly accounted for.

\subsubsection*{Miscalibration}  
We simulate a relative calibration error, uncorrelated between the
channel input maps and applied directly at the map level (so that no
leakage between the Stokes parameters occurs). Though this is clearly
a simplification, we note that it is not completely unrealistic and
may be expected for experiments implementing a fast polarization
modulator, such as a continuously rotating half-wave plate used by
\mbox{MAXIPOL}~\citep{2007ApJ...665...42J} and under development for use with
EBEX~\citep{2008SPIE.7020E..68G}. In such experiments the three Stokes
parameters can be disentangled from data of any single detector, and
the resulting maps co-added a posteriori in a noise-weighted fashion.
The impact of calibration errors and uncertainties can be
mitigated by modelling the calibration parameters at the same time as
estimating the spectral parameters, as mentioned in
Section~\ref{sec:miramare}.

We simulate several cases in which we impose different pre-defined
miscalibration values, centered on $\omega_i = 1 + \delta\omega_i$,
introducing Gaussian priors on the calibration parameters, centered on
$\omega_i=1$ with width $\delta\omega_i$. We then quantify the
  impact of calibration errors by calculating the ratio between the
  residuals in the basic, perfectly calibrated case, and the
  miscalibrated cases.  We report these ratios in
Table~\ref{tab:miscalib} which shows the effect of
miscalibrating the 250 and the 410 GHz channels with respect to the
150 GHz channel, finding for instance that 5\% calibration errors in
  these two channels leads to foreground residuals that are amplified
  by a factor 3. From Figure~\ref{Cases_ps_balloon}, we can see that this
  enhancement of the foreground spectrum by a factor approximately 10
  would impact adversely on the large angular scale B-mode
  estimation.

\subsubsection*{Spectral mismatch}  

\begin{table}
\begin{center}
\caption{This table reports how a mismatch applied to the input dust
  models, expressed in percentage per channel, increases
  the foreground residuals with respect to the basic case.}
\label{tab:mismatch}
\begin{tabular}{cc}
\hline
\smallskip
\ {\bf Spectral} mismatch & $\displaystyle{\frac{\Delta_{\rm mismatch}}{\Delta_{\rm basic}}}$  \\
\hline
\ $0\%, 0\%, 3\%$ &  $\sim 1.5$  \\
\hline
\ $0\%, 3\%, 0\%$ &  $\sim 2.5$  \\
\hline
\ $0\%, 0\%, 7\%$ &  $\sim 4 $ \\
\hline
\ $0\%, 5\%, 0\%$ & $\sim 6 $  \\
\hline
\end{tabular}
\end{center}
\end{table}

Here we consider situations where a mismatch between the true and 
postulated scaling laws for the dust component is present.
In the studied cases, we use different dust scaling laws in the 
simulations, but during the separation process we always assume the same, 
simple dependence as defined in Equation~(\ref{eq:dust_scal}).
The specific laws used in the simulations are: the Model 8 from 
\citet{1999ApJ...524..867F} and an arbitrary mismatch, where the dust 
amplitudes are changed by some factor from their values as expected in 
the model in Equation~(\ref{eq:dust_scal}).
Model 8 is a two-temperature model with two specific spectral
  indices and two temperatures for the dust. Even if its functional
  form is different from Model 3 of Equation~(\ref{eq:dust_scal}), the
  two models are actually very close in the frequency range from $150$
  to $410$ GHz. Fitting the greybody scaling to these Model 8
  simulations, we found that the final foreground residual was small
  and comparable to the other successful cases already shown.

  This motivated us to investigate cases with a larger spectral
  mismatch in which we inserted some discrete multiplicative factors
  in the dust scaling law used to simulate the frequency channels. We
  progressively broke the assumption perfect knowledge of the dust
  spectral behavior, until the model is too far from the simulation,
  leading to B-mode biases. In Table~\ref{tab:mismatch} we report how
  the results deteriorate, in terms of larger residuals
  $\bd{\Delta}^{\rm CMB}$, for some mismatch choices in different
  channels. The basic result is that mismatches of upwards of 5\% in
  the dust scaling can lead to an enhancement by a factor 6 and
  upwards of the foreground residual level. It is perhaps not
  surprising that effective modelling and subtraction of foregrounds
  using few channels and few free parameters depends on the underlying
  smoothness of the foreground frequency scaling.
  
 We note that this test differs from the miscalibration case discussed earlier,
 as only one of the components amplitude is modified and no prior is used
 in the separation process.

\section{Analysis of the residuals}

\label{sect:anaRes}

In the simulation environment we have the power to control details of
all the aspects of the separation process. In this Section we take the
advantage of this fact and investigate the nature and origin of the
residuals $\bd{\Delta}$ shown so far in the paper. We emphasize that
the considerations presented below do not depend on how the estimate
of the spectral parameters have been obtained, and therefore they
apply more generally than just to the specific parametric ML estimator
considered here. In fact, the analogous reasoning could be applied,
and similar conclusions drawn, in a case of any two step approach in
which first the spectral indices estimates are derived and then the
sky components estimated via Equation~(\ref{eqn:step2amps}). This
includes FastICA \citep[see][and reference therein]{2010MNRAS.402..207B}, 
neural networks \citep{2008Ap&SS.318..195N}, 
and Correlated Component Analysis \citep {2007MNRAS.382.1791B}.
 
As introduced in Equation~(\ref{eqn:step2amps}), the operator we apply
to the data set $\bd{d}$ to recover the component estimates $\bd{s}$
is,
 \begin{eqnarray}
\bd{W\l(\bd{\beta}\r)} = \l(\bd{A}^t\l(\bd{\beta}\r)\,\bd{N}^{-1}\,\bd{A}\l(\bd{\beta}\r)\r)^{-1}\,\bd{A}^t\l(\bd{\beta}\r)\,\bd{N}^{-1}\,,
\label{eqn:sep_oper}
 \end{eqnarray}
where we explicitly highlight the dependence on the recovered dust
spectral index $\bd{\beta}$.  Neglecting the presence of the noise,
the data can be written as $\bd{d}=\bd{A}{\l(\bd{\beta_0}\r)}\bd{s_0}$
(hereafter the subscript $0$ refers to true, i.e., input rather than
estimated quantities), and therefore,
\begin{eqnarray}
\bd{s} = \bd{W}\,\bd{d}\, = \bd{W}\,\bd{A}\l(\bd{\beta_0}\r)\,\bd{s_0} \equiv \bd{Z}\l(\bd{\beta}\r)\,\bd{s_0},
\label{eqn:zMatDef}
 \end{eqnarray}
 and thus the residuals can be written down as,
 \begin{eqnarray}
\bd{\Delta} = \bd{s}\, - \bd{\hat{s}_0} = \l(\bd{Z}\l(\bd{\beta\r)}-\bd{I}\r)\,\bd{s_0},
\label{eqn:resViazMat}
 \end{eqnarray}
 where the last definition of $\bd{\Delta}$ coincides with
 Equation~(\ref{eqn:resDef}) if no noise is considered.  Here
 $\bd{\hat{s}_0}$ refers to the true, input component, which is
 modelled in the separation process and is thus a subvector of
 $\bd{s_0}$. The matrix $\bd{I}$ is made of a square unit matrix,
 corresponding to all modelled components, supplemented by extra
 columns of zeros -- one for each unmodelled component.
  
 The size of the matrix, $\bd{Z}$, depend on the number of actual and
 derived sky components and not on the number of the observed
 frequency channels.  In most of the examples shown in this paper,
 $\bd{Z}$ is a $2\times3$ matrix, since we attempt to recover only two
 (out of three) components.  We note that once we have estimated
 $\bd{\beta}$, the operator $\bd{Z}\l(\bd{\beta}\r)$, that transforms
 the input components into the output ones, can be readily computed
 since, in the simulations, we know the input scalings. In this case,
 Equation~(\ref{eqn:resViazMat}) provides insight into the origin of
 the residuals and their relative amplitudes.

We first observe that, 
\begin{eqnarray}
\bd{Z}\l({\bd \beta} ={\bd \beta_0}\r) = \bd{1}, 
\label{eqn:zOfbeta0}
\end{eqnarray}
if the number of assumed and actual components is the same. If there
are more components used for the simulations then subsequently
recovered, this will no longer be the case. However even then the
maximal square block of the matrix $\bd{Z}\l({\bd \beta} ={\bd
  \beta_0}\r)$ will be equal to a unit matrix. This is shown in the
upper part of Table~\ref{tab:z_basic_balloon}. Clearly, even perfect
knowledge of the true dust spectral index does not assure the lack of
the residual. Nevertheless, in such a case each of the recovered
components contains only a contribution of this component plus one due
to the unmodelled signal.  In a specific case of the balloon-borne
experiment considered here nearly all of the unmodelled synchrotron is
added mostly to the recovered CMB signal, given the similar scaling of
both these components in the respective frequency bands.

The matrix $\bd{Z(\beta)}$ computed in a more general and realistic
case, when the spectral index is unknown and needs to be estimated
from the data, is shown in the middle part of
Table~\ref{tab:z_basic_balloon}.  We note first that as before, and
for the same reason, the unmodelled synchrotron contributes
predominantly to the CMB residuals.  Nevertheless, the dust is now
divided in between the two recovered components, though the dust
signal found in the recovered CMB is very subdominant. Also in the
previous case, the recovered CMB component contains the entire CMB
signal, which is completely absent in the recovered dust
template. This is reminiscent of Equation~(\ref{eqn:zOfbeta0}), which
in the present case holds only for the CMB component reflecting the
fact that the perfect black-body derivative scaling is assumed on both
the simulation and recovery stages. In this case again, the CMB
residuals, $\bd{\Delta}^{\rm CMB}$, do not contain any CMB.  This is
no longer true if the calibration errors are allowed for as shown
in the bottom of Table~\ref{tab:z_basic_balloon}. In this case the
recovered CMB component contains only part of the total CMB signal as
determined by $W$-matrix weighted average of the relative calibration
errors for each of the channels. The remainder of the CMB is then
found in the recovered dust. We note however that though for the
calibration errors considered here these effects are small, the
recovered CMB residual is now indeed typically a mixture of the CMB,
dust and synchrotron signals. We point out that, maybe somewhat
counterintuitively, the elements of $\bd{Z}$ in any column do not have
to sum up to unity. This reflects the fact that due to our wrong
assumptions about the spectral parameters values, the estimated
components contain overall `more' of the input components than there
really is. This effect is small, if the estimated $\bd{\beta}$ values
are close to the true ones, as to first order in $\bd{\beta}$, the
columns of $\bd{Z}$ do sum up to (nearly) unity.

\begin{figure}
\begin{center}
  \includegraphics[width=7cm]{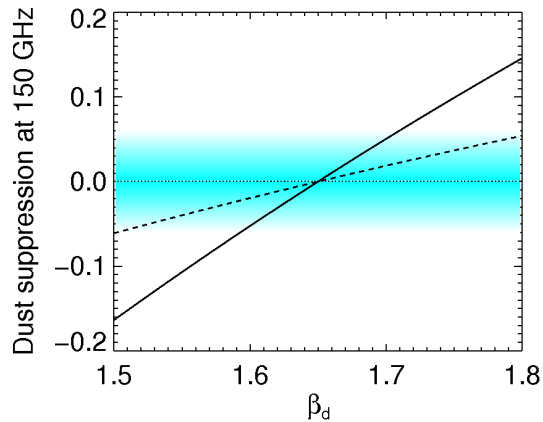}
  \caption{Dust suppression factor in the 150 GHz channel of the basic
      case of the the balloon-borne (solid line) and ground based
      (dashed line) experiments, as a function of the
      recovered dust spectral index $\beta_{\rm d}$. The horizontal
      shaded band is indicative of the requirement on the dust suppression
      factor in order to do B-mode science.}
\label{wa_beta}
\end{center}
\end{figure}

Finally, in all the cases we find that although the code leaves in most
of the unmodelled synchrotron, it cleans the dust to better than $\sim
0.5\%$. The latter depends on the specific value of $\bd{\beta}$
assumed for the recovery as shown in Figure~\ref{wa_beta}, which
shows the relative contamination of the recovered CMB due to the
dust as a function of the recovered spectral index in the case of the
balloon-borne and ground-based basic cases. The shaded band indicates
 the precision which is
needed to avoid contamination in the cosmological B-mode recovery.

The results obtained in the ground-based cases are qualitatively
similar to the ones described above. However, in the basic case, due
to the different assumed frequency coverage, even for the true value
of $\bd{\beta}$ we find non-zero contributions of the synchrotron in
both recovered components. Moreover, the contribution to the CMB is
more than three times that of the actual synchrotron signal at $150$
GHz. Assuming in turn the best-fit value $\bd{\beta}$, we find that
the synchrotron levels in both recovered components remain essentially
unchanged, however the {\it absolute} dust contribution to the CMB 
template increases to become of the same order as that of synchrotron.

\begin{table}
\begin{center}
\caption{This table reports how the input components are weighted in
  the outputs calculated using $\bd{Z}$ as defined in
    Equation~(\ref{eqn:zMatDef}), for the balloon-borne basic case and
  an ideal case where the dust spectral index is known. The RMS values
  of the CMB, dust and synchrotron signal in the studied patch are
  3.1, 0.6, 0.02 $\mu$K, for CMB, dust and synchrotron respectively.}
\label{tab:z_basic_balloon}
\begin{tabular}{cccc}
\hline
\multicolumn{4}{c}{\bf Balloon-borne: Ideal case, $\beta=\beta{_0}=1.65$}\\
\hline
Input:\hfill\hfill & CMB & Dust & Synchrotron  \\ 
\hline
Output:\hfill\hfill & & & \\
CMB & $1.000$   & $0.000$ & $ 1.003$\\
Dust & $0.000$  & $1.000$ & $-0.037$ \\
\hline
\hline
\multicolumn{4}{c}{\bf Basic case, $\beta=1.655$}\\
\hline
Input:\hfill\hfill & CMB & Dust & Synchrotron  \\ 
\hline
Output:\hfill\hfill & & & \\
CMB & $1.000$   & $0.005$ & $1.003$\\
Dust  & $0.000$  & $0.994$ & $-0.036$\\
\hline
\hline
\multicolumn{4}{c}{\bf $5\%$ Miscalibration case, $\beta=1.639$}\\
\hline
Input:\hfill\hfill & CMB & Dust & Synchrotron  \\ 
\hline
Output:\hfill\hfill & & & \\
CMB & $0.988$   & $0.005$ & $0.992$\\
Dust & $-0.0004$  & $0.982$ & $-0.037$\\
\hline
\end{tabular}
\end{center}
\end{table}

\begin{table}
\begin{center}
\caption{This table reports how the input components are weighted in the outputs, for the studied
ground-based cases. The RMS values of the CMB, dust and
  synchrotron signal in the studied patch are 
  3.1, 1.3, 0.02 $\mu$K, for CMB, dust and synchrotron respectively.}
\label{tab:z_ground}
\begin{tabular}{cccc}
\hline
\multicolumn{4}{c}{\bf Ground-based: Ideal case, $\beta=\beta{_0}=1.65$}\\
\hline
Input:\hfill\hfill & CMB & Dust & Synchrotron  \\ 
\hline
Output:\hfill\hfill & & & \\
CMB & $1.000$   & $0.000$ & $ 3.116$\\
Dust & $0.000$  & $1.000$ & $-1.45$ \\
\hline
\hline
\multicolumn{4}{c}{\bf Basic case, $\beta=1.875$}\\
\hline
Input:\hfill\hfill & CMB & Dust & Synchrotron  \\ 
\hline
Output:\hfill\hfill & & & \\
CMB & $1.000$   & $0.067$ & $2.993$\\
Dust  & $0.000$  & $0.919$ & $-1.450$\\
\hline
\hline
\multicolumn{4}{c}{\bf Synchrotron template, $\beta=1.682$}\\
\hline
Input:\hfill\hfill & CMB & Dust & Synchrotron  \\ 
\hline
Output:\hfill\hfill & & & \\
CMB & $1.000$   & $0.004$ & $3.108$\\
Dust & $0.000$  & $0.995$ & $-1.441$\\
\hline
\end{tabular}
\end{center}
\end{table}

\section{Cosmological B-mode detection}
\label{sec:tensor2scalar}

The central question of this work is to understand how much the
recovery of the cosmological B-modes is affected by the presence of
the foregrounds or foreground residuals left over by some foreground
cleaning technique.  This question is often phrased as a question
about the detectable values of the tensor-to-scalar ratio, $r$, of the
corresponding primordial power spectra, and the calculated limiting
values of $r$ are dependent on the implicit and explicit assumptions
made in the course of its derivation.

In the context of this paper this question is, however, well defined
as we study specific foreground separation and power spectrum
estimation algorithms.  Our goal here is to derive values of $r$,
which can not only be detected by the considered experiments but also
convincingly argued for from a perspective of an observer as indeed
being driven by the primordial signal of the cosmological origin.
The limits we aim for here are not to be considered as some `ultimate' lower
limits on $r$, as often quoted in the literature, but rather
representative of the potential of the specific experiments and data
analysis techniques considered here. Reaching values of $r$ lower than
our estimates by the actual experiments once they are deployed and
operating, may not only be plausible but indeed is expected owing to
the build up, over time, of our knowledge of foreground properties and
data analysis tools. 
   
We start by developing a model to describe statistically the
foreground residuals found in the recovered CMB map, thus extending the
pixel-domain discussion of the last section into the power spectrum
domain.  We assume that our estimates of $\bd{\beta}$ are unbiased and
that the obtained statistical uncertainty of the $\bd{\beta}$ recovery
is small, as is indeed the case in the cases considered. Now, if
all the actual components are included in our data model, as defined
by $\bd{W}\l(\bd{\beta}\r)$, it is accurate to write,
\begin{eqnarray}
\bd{Z}\l(\bd{\beta}\r) - \bd{1} & \simeq &\delta\bd{\beta}  \, \frac{\partial\,Z\l(\bd{\beta_0}\r)}{\partial \bd {\beta}},
\end{eqnarray}
where $\delta\bd{\beta}(\equiv\bd{\beta}-\bd{\beta_0})$ is assumed to
be a Gaussian variable with a zero mean and the dispersion as derived
earlier, and we used the fact that $\bd{Z}\l(\bd{\beta_0}\r)=1$, as in
Equation~(\ref{eqn:zOfbeta0}).  Using Equation~(\ref{eqn:resViazMat})
we can then express the residuals of all the modelled components as,
\begin{eqnarray}
\bd{\Delta^{\rm mod}} & = & \l[\bd{Z}\l(\bd{\beta}\r) - \bd{1}\r]\,\bd{s_0} \simeq \delta\bd{\beta}\,\frac{\partial\,\bd{Z}\l(\bd{\beta_0}\r)}{\partial \bd {\beta}}\,\bd{s_0}\\
&= & \delta\bd{\beta}\,\l[\frac{\partial\,\bd{W}\l(\bd{\beta_0}\r)}{\partial \bd {\beta}}\r]\,\bd{A}\l(\bd{\beta_0}\r)\,\bd{s_0},
\label{eqn:delta_est}
\end{eqnarray}
and use a first row of this matrix equation to compute the foreground residuals as found in the recovered CMB map.
We first introduce a tensor, $\bd{\alpha}^{ij}_k$, defined as,
\begin{eqnarray}
\bd{\alpha}^{ij}_k \equiv \frac{\partial\,\bd{Z}_{ij}\l(\bd{\beta_0}\r)}{\partial \bd {\beta}_k}
\end{eqnarray}
and we can then express a combined residual  in the CMB map ($i=0$) due to all the modelled, non-CMB component as,
\begin{eqnarray}
\Delta^{\rm CMB} = \sum_{k,j}\,\delta\bd{\beta}_k\,\bd{\alpha}^{0j}_k \, \bd{s_0}^j.
\end{eqnarray}
This shows that the residuals behave like templates with amplitudes
randomized due to the impact of the CMB and noise variance on the
determination of the spectral parameters. Hereafter we will assume
that the latter are Gaussian variables centered at the true value,
$\bd{\beta_0}$, and with dispersions as estimated from the data,
$\bd{\Sigma}^2(\equiv\l\langle\delta
\bd{\beta}\,\delta\bd{\beta}^t\r\rangle)$.  Consequently, for pure
spectra averaged over the statistical ensemble (of the noise and CMB)
we can then write,
\begin{eqnarray}
\bd{C}^{\Delta}_\ell = \sum_{k,k'}\sum_{j,j'}\,\bd{\Sigma}_{kk'}^2\,\bd{\alpha}^{0j}_k\,\bd{\alpha}^{0j'}_{k'} \, \bd{C}^{jj'}_{\bd{0},\,\ell},
\label{eqn:resSpec0}
\end{eqnarray}
where $\bd{C}^{jj'}_{\bd{0},\,\ell}$ are the (pure) auto- and cross- spectra for every pair of the actual $j$ and $j'$ non-CMB components.
 
 In a case of foreground components which can not be, and/or are not
 modelled in the separation process, we can no longer use the
 procedure outlined above to estimate their residuals.  From the
 previous section we note that such residuals depend only weakly on
 the assumed spectral parameters and thus will not be stochastic over
 the CMB plus noise realizations and should be rather treated as a
 bias.  Moreover, in the specific cases considered in this paper we
 note that the element of the matrix, $\bd{Z}$, which determines the
 weight with which the component is added to the cleaned CMB map is on
 order of at most a few, what together with the fact that the overall
 expected level of the synchrotron is very low, leads to the
 conclusion that indeed it can be neglected for the estimations of $r$
 as derived here. This last statement can be phrased somewhat more
 formally by using the approach of \citet{2007PhRvD..75h3508A}, which
 allows to quantify values of $r$, denoted as $r_{res}$, which will not be affected by a
 bias due to some present residual. Indeed, we define the two
 quantities:
\begin{eqnarray}
s(r) &=& \sum C_l^{cmb}(r),
\\
u &=& \sum C_l^{res}.
\label{eq:amblard}
\end{eqnarray}
In the successful cases considered in the paper we find that typically
$u \sim s(r_{res})$ for $r^{\l(b\r)}_{res}=0.005$ and $r^{\l(g\r)}_{res}=0.015$
for the balloon-borne and ground-based experiments respectively. They represent the smallest
values of $r$ for which the unmodelled residuals due to synchrotron can
be neglected.   

We will now assume that the bias due to the unmodelled components is
negligible and proceed to the estimation of the parameter $r$
accounting for  the extra variance due to the modelled component
residuals. For this we use a Fisher like approach, which in our case
reads, 
\begin{eqnarray}
F(r) = \frac{\partial \bd{C^{\rm CMB}}_b}{\partial r} \,\,\bd{\Sigma}_{bb'}^{-1}\l(r\r)\,\,\frac{\partial \bd{C^{\rm CMB}}_{b'}}{\partial r}
\label{eqn:fisher}
\end{eqnarray}
where $b$ denotes the $\ell$-bin number used for the power spectrum estimation, and,
\begin{eqnarray}
\bd{\Sigma}_{bb'}\l(r\r) & \equiv & {\rm Var}\l({\bd C}^{{\rm CMB+noise}}_{bb'}\r) + \bd{C}^{\rm CMB+noise}_b\,\bd{C^{\Delta}}_{b'} \nonumber \\
                                      & + & \bd{C}^{\rm CMB+noise}_{b'}\,\bd{C^{\Delta}}_{b}  + \bd{C^{\Delta}}_{b} \bd{C^{\Delta}}_{b'}.
\label{eqn:sigma_fisher}
\end{eqnarray}
Here the first term is the covariance of estimated pure CMB B-mode
spectra, which we assume to be diagonal in the bin-space and which is
estimated via MC simulations for a grid of values of
$r$. $\bd{C^{\rm CMB+noise}}$ are the estimated pure CMB+noise spectra
also computed on a grid of $r$ values. The partial derivatives are
computed using the binned theoretical spectra, obtained in this case
from the {\sc camb} code. This is justified given that the pure
estimator is assumed to be unbiased. The last term in
Equation~(\ref{eqn:sigma_fisher}) describes the variance due to the
residual foreground treated here as a template fully correlated
between different bins. The third and fourth terms reflect the
cross-terms between the foreground residuals and CMB+noise. These
again are non-diagonal in the bin domain. We note that the off-diagonal
template-like correction appearing in Equation~(\ref{eqn:sigma_fisher}) corresponds 
roughly to the tensor-product term found in Equation~(A4) of~\citet{2009MNRAS.392..216S}, 
which indeed describes the correction to the noise correlation matrix, 
Equation~(\ref{eqn:noiseCorrOptim}), due to the foreground residuals.

We then search for the values of $r_d$ such that $r \geq r_d \simeq
2\, F^{-1/2}\l(r\r)$, which can hence be detected at a confidence level not smaller
than $95$\%.  For the experimental setups described in this work, we
find that $r_d \sim 0.04$ both for the balloon-borne and the ground-based cases.
This value of $r$ is sufficiently high that the bias expected due to the unmodelled
synchrotron is indeed much smaller than the detected values. We also
find that the derived value of $r$ is limited by the CMB+noise variance, with
the foregrounds effects being subdominant.  This observation is
perhaps not surprising given the low level of the residuals as
discussed in the previous Sections. However, only an analysis as the
one presented in this Section, can properly account for the bin-bin
correlations of the foreground residual templates.

We note that up to this point we have assumed that we are privy to
some insights as to the true nature and morphology of the foreground
signals and their scaling well beyond what is usually available to an
actual CMB observer and data analyst.  In a real life situation we
will lack some of that information. Specifically we will be likely
ignorant of the true value of the $\bd{\beta}$ parameters, i.e.,
$\bd{\beta_0}$, and cross-spectra of all the actual sky components,
$\bd{C}_{\bd{0},\,\ell}^{j,j'}$.  That may not look like a big issue
given our conclusion above stating that the dominant uncertainty will
be due to the CMB and noise variance. However, this conclusion
may need to be corroborated case-by-case using data, analyzed
together with some necessary external information. Whenever
it turns out not to hold then a consistent procedure to account
for the extra effects may be needed.

\begin{figure}
\centering
\includegraphics[width=7.5cm]{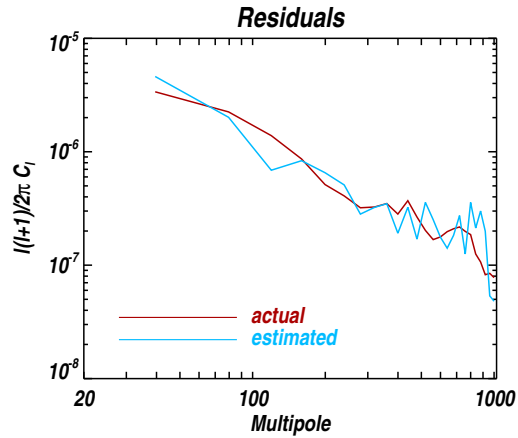}
\caption{Comparison between actual and estimated residuals for the
  ground-based, no synchrotron case, calculated using Equation~(\ref{eqn:resSpec0}).}
\label{resid_comp}
\end{figure}

In the case at hand this can be done by replacing the needed
information with their best estimates derived in the analysis process.
We will thus use the best-fit value in place of $\bd{\beta_0}$ and the
pure spectra of the components estimated as a result of the separation
process, and corrected for the noise bias using
Equation~(\ref{eqn:noiseCorrOptim}) in the case of the auto-spectra,
$j=j'$. In Figure~\ref{resid_comp} we show how this procedure fares in
the case of the dust component in one of the ground-based case example
considered in this paper.  To evaluate the bias due to the unmodelled
synchrotron, we need to use external data. We could correct for such a
bias, if the available data are of sufficiently high precision, and
include the resulting variance in the final power spectrum error
budget. 
However, more typically, one
would rather aim to show that the bias is indeed negligible in the 
sense defined above in Equation~(\ref{eq:amblard}).  For
this task the required external data however need not being very
precise, not least due the fact that the unmodelled components
considered here are assumed to be truly subdominant. 
 
Using the estimated quantities in our $r$-estimation procedure we
recover the limits on $r$ essentially identical to those found before.
We thus conclude that  values of $r\simgt$0.04  are not only 
{\em detectable} at greater than the $95$\%
confidence level, but can be detected in the realm of an actual experiment and
argued for as of a cosmological origin, all of that providing a sufficient control of
systematic effects.

\section{Conclusions}
In this paper we study the performance of the maximum likelihood
parametric component separation method from the point of view of its
application to the CMB B-mode polarization analysis. We investigate
the residuals left over from the separation in both
the pixel and harmonic domains. We propose an efficient
framework for evaluating the pixel domain residuals in the simulation,
and show how it can be used to gain important insights into the
separation process. We then compute the power spectra of the recovered
CMB maps, as well as maps of the residuals, using the pure pseudo
spectra technique, and estimate their variances using Monte Carlo
simulations. Finally, we propose a Fisher-like approach to evaluate
the effects of the foreground residuals on the $r$ parameter and use
the latter to derive some estimates of typical values of $r$, which
are potentially detectable by the considered experiments at the $95$\%
confidence level. The latter estimates thus include the
uncertainties due to sampling variance, noise scatter, E-to-B leakage,
and foreground residuals, all of which are consistently propagated
through the proposed pipeline.

We focus here on the small-scale, bolometric experiments, broadly
  dividing them into two classes, referred to as balloon-borne
and ground-based setups, both observing in three
different frequency bands. We find that the balloon-borne case, with
frequency bands at $150$, $250$, and $410$ GHz, provides a
robust experimental setup for the detection
of the B-mode polarization. The foreground residuals in the recovered CMB maps
derived in this case are found to be usually subdominant. This is true
whenever the assumed data model is indeed correct, but it also holds
when some small systematic effects are permitted. Selected effects
of this kind considered in this work include relative
calibration errors, unmodelled spatial variation of the spectral
parameters, and spectral mismatch between assumed and true spectral
scaling laws.  We emphasize that all these systematics, though
manageable if sufficiently small, may lead to spurious effects in
general, and therefore need to be controlled in actual experiments
with a sufficient precision, which need to be determined specifically
for any experiment.

The success of the considered balloon-borne cases is related to the
wide frequency range available to such experiments, which permits
selecting frequency bands at the sufficiently high frequencies to
avoid the unwanted residual synchrotron. The latter is found to be a
dominant source of the bias for the ground-based experiments. In the balloon
case we can also afford a long leverage arm between the lowest
(CMB-dominated) and the highest (dust-dominated) frequency bands,
which plays a pivotal role in setting tight constraints on the
spectral parameters of the dust. From our Fisher-like analysis, we show 
that one could detect $r$ values as low as 
$0.04$ at the $95\%$ confidence level with such experiments, if both
our models and measurements are sufficiently well characterized.

For the ground-based case the atmospheric loading restricting the
available frequency window proves to be a significant limitation. We
find that even in an absence of any systematic effects with three
frequency bands set at $90$, $150$, and $220$ GHz, it is generally not
possible to produce sufficiently clean CMB maps. This is due to the
unmodelled, and thus not separated, synchrotron contribution, which is
significant enough (if the polarized emission is at the level
suggested by WMAP) at these frequencies to bias the estimation
of the dust spectral parameters. We point out that this contribution
has been neglected in some earlier works, which consequently has
arrived at a different conclusion. This  therefore emphasizes the
  importance of accurate sky modelling for this kind of the analysis.
Nevertheless, we find a satisfactory cleaning can be achieved in such
a case if some external information is available. In particular, we
discuss the extended ground experiment analysis allowing for the presence of the
extra lower frequency channels, rough synchrotron templates, and
priors on the dust scalings. We find that in such cases, and under
realistic assumptions, the ground-based experiments should reach
a sensitivity roughly matching those found in the balloon-borne
case, in terms of a detectable $r$ parameter. We also conclude that,
for both these types of the experiments, the foreground
contamination anticipated in low-contrast dust regions,  should not  
be an obstacle preventing them from exploring the parameter space of $r$ down to
the values $\sim 0.04$. Indeed, for the considered experimental setups this limit 
is determined by the uncertainty due to the CMB itself and the instrumental 
noise, with the effects of the residual foregrounds found to be sub-dominant. 
In the realm of the actual observations, whether these limits are reached 
will be crucially dependent on the control of systematic effects. We note 
here however that the limits derived on $r$ are strongly dependent
on the level of the noise assumed in the input single-channel maps. These
can be therefore improved upon, if a deeper integration of the same field
is performed. However, if no additional external information is used, those
limits will remain appropriately higher than the $r_{res}$ values obtained 
earlier, and below which foreground bias would become significant.

In this context we point out that the framework described in this paper 
provides a blueprint for similar studies focused this time on systematic 
effects. It can be also extended to perform a realistic experiment 
optimization procedure from the viewpoint of detection of the B-mode signal of 
cosmological origin.

\section*{Acknowledgments}
We thank A. Lee and S. Hanany for helpful comments on the manuscript.
 This research used resources of the National Energy Research
 Scientific Computing Center, which is supported by the Office of
 Science of the U.S. Department of Energy under Contract
 No. DE-AC02-05CH11231. Some of the results in this paper have been
 derived using the {\sc HEALPix}
 package~\citep{2005ApJ...622..759G}. We acknowledge the use of the
 {\sc CAMB} \citep{2000ApJ...538..473L} and
 CosmoMC~\citep{2002PhRvD..66j3511L} packages.  We acknowledge the use
 of the Legacy Archive for Microwave Background Data Analysis
 (LAMBDA). Support for LAMBDA is provided by the NASA Office of Space
 Science.  FS and RS acknowledge the partial support of the French National 
Agency
 (ANR) through COSINUS program (project MIDAS no. ANR-09-COSI-009).
We also thank the ANR-MIDAS'09 project team for helpful discussions. JG 
acknowledges financial support from the Groupement d'Int\'er\^et Scientifique
(GIS) `consortium Physique des 2 Infinis (P2I)'. CB and SL have been
partially supported by ASI contract I/016/07/0 "COFIS".

\bibliographystyle{mn2e}
\bibliography{arx}

\end{document}